%% file: main.tex
\pdfoutput=1

\documentclass[12pt,a4paper]{article}

\usepackage{pdflscape}
\usepackage{placeins}
\usepackage{ifthen} 
\newboolean{pdflatex}
\setboolean{pdflatex}{true} 

\newboolean{articletitles}
\setboolean{articletitles}{true} 

\newboolean{uprightparticles}
\setboolean{uprightparticles}{false} 

\newboolean{inbibliography}
\setboolean{inbibliography}{false} 

\input{preamble}

\input{shortcuts}

\begin{document}

\renewcommand{\thefootnote}{\fnsymbol{footnote}}
\setcounter{footnote}{1}

\input{title-LHCb-PAPER}


\renewcommand{\thefootnote}{\arabic{footnote}}
\setcounter{footnote}{0}



\pagestyle{plain} 
\setcounter{page}{1}
\pagenumbering{arabic}


%

\input{introduction}

\input{detector}

\input{selection}

\input{fit-model}

\input{efficiencies}

\input{systematics}

\input{results}

\input{acknowledgements}

\input{appendix}

\addcontentsline{toc}{section}{References}
\setboolean{inbibliography}{true}
\bibliographystyle{LHCb}
\bibliography{main,LHCb-PAPER,LHCb-CONF,LHCb-DP,LHCb-TDR}

 
\newpage
\input{LHCb_Authorship_flat_07-Mar-2017.tex}



\end{document}

%% file: preamble.tex
\usepackage[top=1in, bottom=1.25in, left=1in, right=1in]{geometry}

\usepackage{microtype}
\usepackage{lineno}  
\usepackage{xspace} 
\usepackage{caption} 

\usepackage{graphicx}  
\usepackage{color}
\usepackage{colortbl}
\graphicspath{{./figs/}} 

\usepackage{amsmath} 
\usepackage{amssymb}
\usepackage{amsfonts}
\usepackage{upgreek} 

\newcommand*\patchAmsMathEnvironmentForLineno[1]{%
\expandafter\let\csname old#1\expandafter\endcsname\csname #1\endcsname
\expandafter\let\csname oldend#1\expandafter\endcsname\csname
end#1\endcsname
 \renewenvironment{#1}%
   {\linenomath\csname old#1\endcsname}%
   {\csname oldend#1\endcsname\endlinenomath}%
}
\newcommand*\patchBothAmsMathEnvironmentsForLineno[1]{%
  \patchAmsMathEnvironmentForLineno{#1}%
  \patchAmsMathEnvironmentForLineno{#1*}%
}
\AtBeginDocument{%
\patchBothAmsMathEnvironmentsForLineno{equation}%
\patchBothAmsMathEnvironmentsForLineno{align}%
\patchBothAmsMathEnvironmentsForLineno{flalign}%
\patchBothAmsMathEnvironmentsForLineno{alignat}%
\patchBothAmsMathEnvironmentsForLineno{gather}%
\patchBothAmsMathEnvironmentsForLineno{multline}%
\patchBothAmsMathEnvironmentsForLineno{eqnarray}%
}

\usepackage{hyperref}    
\usepackage[all]{hypcap} 

\input{lhcb-symbols-def} 

\usepackage{cite} 
\usepackage{mciteplus}

%% file: lhcb-symbols-def.tex
\usepackage{xspace} 
\usepackage{upgreek}

\def\lhcb {\mbox{LHCb}\xspace}

\def\babar  {\mbox{BaBar}\xspace}

\def\velo   {VELO\xspace}

\def\MagUp {\mbox{\em Mag\kern -0.05em Up}\xspace}

\ifthenelse{\boolean{uprightparticles}}%
{
 \def\Pbeta       {\ensuremath{\upbeta}\xspace}
 \def\Pgamma      {\ensuremath{\upgamma}\xspace}

 \def\Peta        {\ensuremath{\upeta}\xspace}

 \def\Pmu         {\ensuremath{\upmu}\xspace}

 \def\Ppi         {\ensuremath{\uppi}\xspace}                 
                  
 \def\Prho        {\ensuremath{\uprho}\xspace}

 \def\Pchi        {\ensuremath{\upchi}\xspace}                 
 \def\Ppsi        {\ensuremath{\uppsi}\xspace}

 \def\PDelta      {\ensuremath{\Delta}\xspace}                 
 \def\PXi      {\ensuremath{\Xi}\xspace}                 
 \def\PLambda      {\ensuremath{\Lambda}\xspace}                 
 \def\PSigma      {\ensuremath{\Sigma}\xspace}                 
 \def\POmega      {\ensuremath{\Omega}\xspace}                 
 \def\PUpsilon      {\ensuremath{\Upsilon}\xspace}                 
 
 \def\PB      {\ensuremath{\mathrm{B}}\xspace}                 
                  
 \def\PD      {\ensuremath{\mathrm{D}}\xspace}

 \def\PJ      {\ensuremath{\mathrm{J}}\xspace}                 
 \def\PK      {\ensuremath{\mathrm{K}}\xspace}

 \def\Pb      {\ensuremath{\mathrm{b}}\xspace}                 
 \def\Pc      {\ensuremath{\mathrm{c}}\xspace}                 
 \def\Pd      {\ensuremath{\mathrm{d}}\xspace}

 \def\Ph      {\ensuremath{\mathrm{h}}\xspace}                 
 \def\Pi      {\ensuremath{\mathrm{i}}\xspace}

 \def\Pp      {\ensuremath{\mathrm{p}}\xspace}                 
 \def\Pq      {\ensuremath{\mathrm{q}}\xspace}                 
                  
 \def\Ps      {\ensuremath{\mathrm{s}}\xspace}

}
{
 \def\Pbeta       {\ensuremath{\beta}\xspace}
 \def\Pgamma      {\ensuremath{\gamma}\xspace}

 \def\Peta        {\ensuremath{\eta}\xspace}

 \def\Pmu         {\ensuremath{\mu}\xspace}

 \def\Ppi         {\ensuremath{\pi}\xspace}                 
                  
 \def\Prho        {\ensuremath{\rho}\xspace}

 \def\Pchi        {\ensuremath{\chi}\xspace}                 
 \def\Ppsi        {\ensuremath{\psi}\xspace}                 
                  
 \mathchardef\PDelta="7101
 \mathchardef\PXi="7104
 \mathchardef\PLambda="7103
 \mathchardef\PSigma="7106
 \mathchardef\POmega="710A
 \mathchardef\PUpsilon="7107
                  
 \def\PB      {\ensuremath{B}\xspace}                 
                  
 \def\PD      {\ensuremath{D}\xspace}

 \def\PJ      {\ensuremath{J}\xspace}                 
 \def\PK      {\ensuremath{K}\xspace}

 \def\Pb      {\ensuremath{b}\xspace}                 
 \def\Pc      {\ensuremath{c}\xspace}                 
 \def\Pd      {\ensuremath{d}\xspace}

 \def\Ph      {\ensuremath{h}\xspace}                 
 \def\Pi      {\ensuremath{i}\xspace}

 \def\Pp      {\ensuremath{p}\xspace}                 
 \def\Pq      {\ensuremath{q}\xspace}                 
                  
 \def\Ps      {\ensuremath{s}\xspace}

}

\makeatletter
\ifcase \@ptsize \relax
  \newcommand{\miniscule}{\@setfontsize\miniscule{4}{5}}
\or
  \newcommand{\miniscule}{\@setfontsize\miniscule{5}{6}}
\or
  \newcommand{\miniscule}{\@setfontsize\miniscule{5}{6}}
\fi
\makeatother

\DeclareRobustCommand{\optbar}[1]{\shortstack{{\miniscule (\rule[.5ex]{1.25em}{.18mm})}
  \\ [-.7ex] $#1$}}

\def\mup        {{\ensuremath{\Pmu^+}}\xspace}
\def\mun        {{\ensuremath{\Pmu^-}}\xspace} 

\def\quark     {{\ensuremath{\Pq}}\xspace}
\def\quarkbar  {{\ensuremath{\overline \quark}}\xspace}

\def\dquark    {{\ensuremath{\Pd}}\xspace}

\def\squark    {{\ensuremath{\Ps}}\xspace}

\def\cquark    {{\ensuremath{\Pc}}\xspace}
\def\cquarkbar {{\ensuremath{\overline \cquark}}\xspace}

\def\bquark    {{\ensuremath{\Pb}}\xspace}

\def\pion   {{\ensuremath{\Ppi}}\xspace}
\def\piz    {{\ensuremath{\pion^0}}\xspace}

\def\pip    {{\ensuremath{\pion^+}}\xspace}
\def\pim    {{\ensuremath{\pion^-}}\xspace}

\def\pimp   {{\ensuremath{\pion^\mp}}\xspace}

\def\rhoz     {{\ensuremath{\rhomeson^0}}\xspace}

\def\kaon    {{\ensuremath{\PK}}\xspace}
  \def\Kbar    {{\kern 0.2em\overline{\kern -0.2em \PK}{}}\xspace}

\def\KorKbar    {\kern 0.18em\optbar{\kern -0.18em K}{}\xspace}
\def\Kz      {{\ensuremath{\kaon^0}}\xspace}

\def\Kp      {{\ensuremath{\kaon^+}}\xspace}
\def\Km      {{\ensuremath{\kaon^-}}\xspace}
\def\Kpm     {{\ensuremath{\kaon^\pm}}\xspace}

\def\KS      {{\ensuremath{\kaon^0_{\mathrm{ \scriptscriptstyle S}}}}\xspace}

\def\Kstarz  {{\ensuremath{\kaon^{*0}}}\xspace}
\def\Kstarzb {{\ensuremath{\Kbar{}^{*0}}}\xspace}

\newcommand{\etapr}{\ensuremath{\Peta^{\prime}}\xspace}

  \def\Dbar    {{\kern 0.2em\overline{\kern -0.2em \PD}{}}\xspace}
\def\D       {{\ensuremath{\PD}}\xspace}

\def\DorDbar    {\kern 0.18em\optbar{\kern -0.18em D}{}\xspace}
\def\Dz      {{\ensuremath{\D^0}}\xspace}

\def\Dp      {{\ensuremath{\D^+}}\xspace}

\def\Dstarp  {{\ensuremath{\D^{*+}}}\xspace}

\def\Dsp     {{\ensuremath{\D^+_\squark}}\xspace}

\def\B       {{\ensuremath{\PB}}\xspace}
\def\Bbar    {{\ensuremath{\kern 0.18em\overline{\kern -0.18em \PB}{}}}\xspace}

\def\BorBbar    {\kern 0.18em\optbar{\kern -0.18em B}{}\xspace}
\def\Bz      {{\ensuremath{\B^0}}\xspace}

\def\Bd      {{\ensuremath{\B^0}}\xspace}
\def\Bs      {{\ensuremath{\B^0_\squark}}\xspace}

\def\jpsi     {{\ensuremath{{\PJ\mskip -3mu/\mskip -2mu\Ppsi\mskip 2mu}}}\xspace}

\def\chiczero {{\ensuremath{\Pchi_{\cquark 0}}}\xspace}

  \def\Y#1S{\ensuremath{\PUpsilon{(#1S)}}\xspace}

\def\proton      {{\ensuremath{\Pp}}\xspace}

\def\Lz          {{\ensuremath{\PLambda}}\xspace}
\def\Lbar        {{\ensuremath{\kern 0.1em\overline{\kern -0.1em\PLambda}}}\xspace}
\def\LorLbar    {\kern 0.18em\optbar{\kern -0.18em \PLambda}{}\xspace}

\def\Lb      {{\ensuremath{\Lz^0_\bquark}}\xspace}

\def\Lc      {{\ensuremath{\Lz^+_\cquark}}\xspace}

\def\BF         {{\ensuremath{\mathcal{B}}}\xspace}

\def\BR         {\BF}
\newcommand{\decay}[2]{\ensuremath{#1\!\to #2}\xspace}         

\def\to                 {\ensuremath{\rightarrow}\xspace}

\def\order   {{\ensuremath{\mathcal{O}}}\xspace}

\def\eps   {{\ensuremath{\varepsilon}}\xspace}

\def\CP                {{\ensuremath{C\!P}}\xspace}

\newcommand{\DGs}{{\ensuremath{\Delta\Gamma_{\squark}}}\xspace}

\newcommand{\Gs}{{\ensuremath{\Gamma_{\squark}}}\xspace}

\def\AT#1     {\ensuremath{A_{\mathrm{T}}^{#1}}\xspace}           

\def\C#1      {\ensuremath{\mathcal{C}_{#1}}\xspace}                       
\def\Cp#1     {\ensuremath{\mathcal{C}_{#1}^{'}}\xspace}                    
\def\Ceff#1   {\ensuremath{\mathcal{C}_{#1}^{\mathrm{(eff)}}}\xspace}        
\def\Cpeff#1  {\ensuremath{\mathcal{C}_{#1}^{'\mathrm{(eff)}}}\xspace}       
\def\Ope#1    {\ensuremath{\mathcal{O}_{#1}}\xspace}                       
\def\Opep#1   {\ensuremath{\mathcal{O}_{#1}^{'}}\xspace}                    

\newcommand{\tev}{\ifthenelse{\boolean{inbibliography}}{\ensuremath{~T\kern -0.05em eV}\xspace}{\ensuremath{\mathrm{\,Te\kern -0.1em V}}}\xspace}
\newcommand{\gev}{\ensuremath{\mathrm{\,Ge\kern -0.1em V}}\xspace}
\newcommand{\mev}{\ensuremath{\mathrm{\,Me\kern -0.1em V}}\xspace}
\newcommand{\kev}{\ensuremath{\mathrm{\,ke\kern -0.1em V}}\xspace}
\newcommand{\ev}{\ensuremath{\mathrm{\,e\kern -0.1em V}}\xspace}
\newcommand{\gevc}{\ensuremath{{\mathrm{\,Ge\kern -0.1em V\!/}c}}\xspace}
\newcommand{\mevc}{\ensuremath{{\mathrm{\,Me\kern -0.1em V\!/}c}}\xspace}
\newcommand{\gevcc}{\ensuremath{{\mathrm{\,Ge\kern -0.1em V\!/}c^2}}\xspace}
\newcommand{\gevgevcccc}{\ensuremath{{\mathrm{\,Ge\kern -0.1em V^2\!/}c^4}}\xspace}
\newcommand{\mevcc}{\ensuremath{{\mathrm{\,Me\kern -0.1em V\!/}c^2}}\xspace}

\def\mm   {\ensuremath{\mathrm{ \,mm}}\xspace}

\def\mum  {\ensuremath{{\,\upmu\mathrm{m}}}\xspace}

\def\invfb   {\ensuremath{\mbox{\,fb}^{-1}}\xspace}

\newcommand{\stat}{\ensuremath{\mathrm{\,(stat)}}\xspace}
\newcommand{\syst}{\ensuremath{\mathrm{\,(syst)}}\xspace}

\def\order{{\ensuremath{\mathcal{O}}}\xspace}
\newcommand{\chisq}{\ensuremath{\chi^2}\xspace}

\newcommand{\chisqip}{\ensuremath{\chi^2_{\text{IP}}}\xspace}
\newcommand{\chisqvs}{\ensuremath{\chi^2_{\text{VS}}}\xspace}
\newcommand{\chisqvtx}{\ensuremath{\chi^2_{\text{vtx}}}\xspace}

\def\gsim{{~\raise.15em\hbox{$>$}\kern-.85em
          \lower.35em\hbox{$\sim$}~}\xspace}
\def\lsim{{~\raise.15em\hbox{$<$}\kern-.85em
          \lower.35em\hbox{$\sim$}~}\xspace}

\def\sPlot{\mbox{\em sPlot}\xspace}

\def\ptot       {\mbox{$p$}\xspace}
\def\pt         {\mbox{$p_{\mathrm{ T}}$}\xspace}

\def\evtgen     {\mbox{\textsc{EvtGen}}\xspace}

\def\geant      {\mbox{\textsc{Geant4}}\xspace}

\def\photos     {\mbox{\textsc{Photos}}\xspace}

\def\pythia     {\mbox{\textsc{Pythia}}\xspace}

\def\tell1  {TELL1\xspace}
\def\ukl1   {UKL1\xspace}



%% file: shortcuts.tex
\ifthenelse{\boolean{uprightparticles}}%
{\def\Prho      {\ensuremath{\uprho}\xspace}
 
}
{\def\Prho      {\ensuremath{\rho}\xspace}
 
}
\def\rhoz   {\ensuremath{\Prho^0}\xspace}

\def\had  {\ensuremath{\Ph}\xspace}
\def\hadp  {\ensuremath{\Ph^+}\xspace}
\def\hadm  {\ensuremath{\Ph^-}\xspace}
\def\hadpm  {\ensuremath{\Ph^{\pm}}\xspace}

\def\hadprim  {\ensuremath{\Ph^{\prime}}\xspace}

\def\hadprimm {\ensuremath{\Ph^{\prime-}}\xspace}

\def\hadprimmp  {\ensuremath{\Ph^{\prime\mp}}\xspace}

\def\Bdsz      {\ensuremath{\B^0_{(s)}}\xspace}

\def\BdorBdbar    {\ensuremath{\kern 0.18em\optbar{\kern -0.18em B}{}^0}\xspace}
\def\BsorBsbar    {\ensuremath{\kern 0.18em\optbar{\kern  0.06em B_s}{}^0}\xspace}

\def\pipi  {\ensuremath{\pion^+\pion^-}\xspace}

\newcommand{\subdecay}[2]{\ensuremath{#1(\!\to #2)}\xspace}
\def\BdtoKzKK   {\decay{\Bd}{\Kz \Kp \Km}}
\def\BdtoKzPiPi   {\decay{\Bd}{\Kz \pip \pim}}
\def\BdtoKzKPi   {\decay{\Bd}{\KorKbar^0 \Kpm \pimp}}

\def\BstoKzKK   {\decay{\Bs}{\Kz \Kp \Km}}
\def\BstoKzPiPi   {\decay{\Bs}{\Kz \pip \pim}}
\def\BstoKzKPi   {\decay{\Bs}{\KorKbar^0 \Kpm \pimp}}

\def\BdtoKsKK   {\decay{\Bd}{\KS \Kp \Km}}
\def\BdtoKsPiPi   {\decay{\Bd}{\KS \pip \pim}}
\def\BdtoKsKPi   {\decay{\Bd}{\KS \Kpm \pimp}}
\def\BdtoKsKpPim   {\decay{\Bd}{\KS \Kp \pim}}
\def\BdtoKsPipKm   {\decay{\Bd}{\KS \Km \pip}}

\def\BstoKsKK   {\decay{\Bs}{\KS \Kp \Km}}
\def\BstoKsPiPi   {\decay{\Bs}{\KS \pip \pim}}
\def\BstoKsKPi   {\decay{\Bs}{\KS \Kpm \pimp}}
\def\BstoKsKpPim   {\decay{\Bs}{\KS \Kp \pim}}
\def\BstoKsPipKm   {\decay{\Bs}{\KS \Km \pip}}

\def\Kshh{\ensuremath{\KS \hadp \hadm}\xspace}
\def\Kshhp{\ensuremath{\KS \hadpm \hadprimmp}\xspace}

\def\KsPiPi{\ensuremath{\KS \pip \pim}\xspace}
\def\KsKPi{\ensuremath{\KS \Kpm \pimp}\xspace}

\def\KsKK{\ensuremath{\KS \Kp \Km}\xspace}

\def\KsKpPim{\ensuremath{\KS \Kp \pim}\xspace}

\def\KsPipKm{\ensuremath{\KS \pip \Km}\xspace}

\def\BdstoKshhp   {\decay{\Bdsz}{\KS \hadpm \hadprimmp}}

\def\BdtoetapKs {\decay{\Bd}{\etapr \KS}}

\def\BstoKstzKstzbtoKsPizKPi  {\decay{\Bs}{\Kstarz (\rightarrow \KS\piz) \Kstarzb (\Km\pip)}}

\def\LbtoKsPip         {\decay{\Lb}{\KS \proton \pim}}

\def\btoqqbars {\decay{\bquark}{\quark\quarkbar\squark}}

\def\btoccbars {\decay{\bquark}{\cquark\cquarkbar\squark}}

\def\LL   {long}
\def\DD   {downstream}

\newcommand{\Br}[1]{\ensuremath{\BF\!\left(#1\right)}\xspace}
\newcommand{\Brr}[1]{\ensuremath{\BF(#1)}\xspace}

\def\fsfdinline {\ensuremath{f_{\squark}/f_{\dquark}}\xspace}

\newcommand{\tab}[1]{Table~\ref{tab : #1}}
\newcommand{\tabs}[2]{Tables~\ref{tab : #1}--\ref{tab : #2}}
\newcommand{\tabstwo}[2]{Tables~\ref{tab : #1} and~\ref{tab : #2}}

\newcommand{\figstwo}[2]{Figs.~\ref{fig : #1} and~\ref{fig : #2}}

\newcommand{\refapp}[1]{Appendix~\ref{sec : #1}}

\def\phz {\phantom{0}}

\newcommand{\plotIfExists}[2]{
  \IfFileExists{#1}{\includegraphics[width=#2\textwidth]{#1}}{\includegraphics[width=#2\textwidth]{#1}}
}    
\newcommand{\plotOne}[2]{
  \ifthenelse{\boolean{pdflatex}}{
    \plotIfExists{#1.pdf}{#2}
  }{
    \plotIfExists{#1.eps}{#2}
  }
}

\newcommand{\plotDataFitResults}[3]{ 
  \begin{figure}[p]
    \begin{center}
      \ifthenelse{\equal{#1}{Loose}}
                 {\def \bdtComment{BDT optimisation corresponding to the favoured decay modes}}
                 {\def \bdtComment{BDT optimisation corresponding to the suppressed modes}}
                 
                 \plotOne{figs/from5150-Louis-Polynomial-StrongPcut-#1-KSKK#2_#3-Standard-DoubleCB-NoPull_log}{0.35}
                 \plotOne{figs/from5150-Louis-Polynomial-StrongPcut-#1-KSKK#2_#3-Standard-DoubleCB-NoPull}{0.35}\\
                 \plotOne{figs/from5150-Louis-Polynomial-StrongPcut-#1-KSKpi#2_#3-Standard-DoubleCB-NoPull_log}{0.35}
                 \plotOne{figs/from5150-Louis-Polynomial-StrongPcut-#1-KSKpi#2_#3-Standard-DoubleCB-NoPull}{0.35}\\
                 \plotOne{figs/from5150-Louis-Polynomial-StrongPcut-#1-KSpiK#2_#3-Standard-DoubleCB-NoPull_log}{0.35}
                 \plotOne{figs/from5150-Louis-Polynomial-StrongPcut-#1-KSpiK#2_#3-Standard-DoubleCB-NoPull}{0.35}\\
                 \plotOne{figs/from5150-Louis-Polynomial-StrongPcut-#1-KSpipi#2_#3-Standard-DoubleCB-NoPull_log}{0.35}
                 \plotOne{figs/from5150-Louis-Polynomial-StrongPcut-#1-KSpipi#2_#3-Standard-DoubleCB-NoPull}{0.35}\\
                 
                 \ifthenelse{\equal{#2}{DD}}{
                   \caption{Results of the simultaneous fit to data (\DD, #3) with the \bdtComment. The modes \KsKK, \KsKpPim, \KsPipKm and \KsPiPi are shown from top to bottom. The left-hand side plots show the results with a logarithmic scale and the right-hand side with a linear scale. Legend is similar to that of plots shown in Fig.~\ref{fig : fitLoose2011}.}
                 }{
                   \caption{Results of the simultaneous fit to data (\LL, #3) with the \bdtComment. The modes \KsKK, \KsKpPim, \KsPipKm and \KsPiPi are shown from top to bottom. The left-hand side plots show the results with a logarithmic scale and the right-hand side with a linear scale. Legend is similar to that of plots shown in Fig.~\ref{fig : fitLoose2011}.}
                 }
                 \label{fig:FitResult:#1:#2:#3}
    \end{center}
  \end{figure}
}
\newcommand{\plotSignalMCFitResults}[3]{
  \begin{figure}[!htbp]
    \begin{center}
      \plotOne{figs/FitModel/Signal/#1/Bd2KSKK#2_#3-MCFit-Louis-DoubleCB-Standard_log}{0.35}
      \plotOne{figs/FitModel/Signal/#1/Bs2KSKK#2_#3-MCFit-Louis-DoubleCB-Standard_log}{0.35}\\
      \plotOne{figs/FitModel/Signal/#1/Bd2KSKpi#2_#3-MCFit-Louis-DoubleCB-Standard_log}{0.35}
      \plotOne{figs/FitModel/Signal/#1/Bs2KSKpi#2_#3-MCFit-Louis-DoubleCB-Standard_log}{0.35}\\
      \plotOne{figs/FitModel/Signal/#1/Bd2KSpiK#2_#3-MCFit-Louis-DoubleCB-Standard_log}{0.35}
      \plotOne{figs/FitModel/Signal/#1/Bs2KSpiK#2_#3-MCFit-Louis-DoubleCB-Standard_log}{0.35}\\
      \plotOne{figs/FitModel/Signal/#1/Bd2KSpipi#2_#3-MCFit-Louis-DoubleCB-Standard_log}{0.35}
      \plotOne{figs/FitModel/Signal/#1/Bs2KSpipi#2_#3-MCFit-Louis-DoubleCB-Standard_log}{0.35}\\
      \ifthenelse{\equal{#2}{DD}}{
        \caption{Result of the simultaneous fit to simulated samples of the signal decays for \DD \KS reconstruction mode, using the \MakeLowercase{#1} optimisation of the BDT, and shown using a logarithmic scale. \KsKK, \KsKpPim, \KsPipKm, and \KsPiPi are shown from top to bottom, while \Bd decays are shown on the left and \Bs decays on the right. }
      }
                 {
                   \caption{Result of the simultaneous fit to simulated samples of the signal decays for \LL \KS reconstruction mode, using the \MakeLowercase{#1} optimisation of the BDT, and shown using a logarithmic scale. \KsKK, \KsKpPim, \KsPipKm, and \KsPiPi are shown from top to bottom, while \Bd decays are shown on the left and \Bs decays on the right. }
                 }
                 \label{fig:FitModel:Signal:#1:#2:#3}
    \end{center}
\end{figure}
}

\newcommand{\plotSplots}[3]{
  \ifthenelse{\equal{#3}{log}}
             {\def\suffix{FitStandard_log}}
             {\def\suffix{FitStandard}}
             \ifthenelse{\equal{#3}{log}}
                        {\def\scale{linear}\xspace}
                        {\def\scale{logarithmic}\xspace}
                        \begin{figure}[!htbp]
                          \begin{center}
                            \plotOne{figs/#1/sWeights-from5150-Louis-PolSlopes-StrongPcut-#1-KSKKDD_#2_\suffix}{0.35}
                            \plotOne{figs/#1/sWeights-from5150-Louis-PolSlopes-StrongPcut-#1-KSKKLL_#2_\suffix}{0.35}\\
                            \plotOne{figs/#1/sWeights-from5150-Louis-PolSlopes-StrongPcut-#1-KSKpiDD_#2_\suffix}{0.35}
                            \plotOne{figs/#1/sWeights-from5150-Louis-PolSlopes-StrongPcut-#1-KSKpiLL_#2_\suffix}{0.35}\\
                            \plotOne{figs/#1/sWeights-from5150-Louis-PolSlopes-StrongPcut-#1-KSpiKDD_#2_\suffix}{0.35}
                            \plotOne{figs/#1/sWeights-from5150-Louis-PolSlopes-StrongPcut-#1-KSpiKLL_#2_\suffix}{0.35}\\
                            \plotOne{figs/#1/sWeights-from5150-Louis-PolSlopes-StrongPcut-#1-KSpipiDD_#2_\suffix}{0.35}
                            \plotOne{figs/#1/sWeights-from5150-Louis-PolSlopes-StrongPcut-#1-KSpipiLL_#2_\suffix}{0.35}\\
                            \caption{Result of the invariant mass fits used for the sWeights extraction with the \MakeLowercase{#1} BDT optimisation on #2 data (\scale scale). \KsKK, \KsKpPim, \KsPipKm, and \KsPiPi are shown from top to bottom, \DD on the left, \LL on the right.}
                            \label{fig:FitResult:Splots:#1:#2:#3}
                          \end{center}
                        \end{figure}
}

\newcommand{\makeSystTabular}[2]{
  \begin{table}[!htbp]
    \begin{center}
      \ifthenelse{\equal{#2}{KK}}
                 {       
                   \caption{Systematic uncertainties originating from the fit model of the \KsKK modes (using \MakeLowercase{#1} BDT optimisation). The numbers are rounded to the upper integer value, except for the total yield. The total systematic error is defined as the sum in quadrature of all the components.}
                 }{}
                 \ifthenelse{\equal{#2}{Kpi}}
                            {
                              \caption{Systematic uncertainties originating from the fit model of the \KsKpPim modes (using \MakeLowercase{#1} BDT optimisation). The numbers are rounded to the upper integer value, except for the total yield. The total systematic error is defined as the sum in quadrature of all the components.}
                            }{}
                            \ifthenelse{\equal{#2}{piK}}        
                                       {
                                         \caption{Systematic uncertainties originating from the fit model of the \KsPipKm modes (using \MakeLowercase{#1} BDT optimisation). The numbers are rounded to the upper integer value, except for the total yield. The total systematic error is defined as the sum in quadrature of all the components.}
                                       }{}
                                       \ifthenelse{\equal{#2}{pipi}}
                                                  {
                                                    \caption{Systematic uncertainties originating from the fit model of the \KsPiPi modes (using \MakeLowercase{#1} BDT optimisation). The numbers are rounded to the upper integer value, except for the total yield. The total systematic error is defined as the sum in quadrature of all the components.}
                                                  }{}
                                                  \label{Table:FitSyst:#1:#2}
                                                  \resizebox{\textwidth}{!}{
                                                    \input{tables/SystI-PolSlopes-from5150-#1-#2.txt}
                                                  }
    \end{center}
  \end{table}
  
}

\newcommand{\makeInvMass}[1]{
  \ifthenelse{\equal{#1}{Bd2KSpipi} \or \equal{#1}{Bs2KSpipi}}              {\def\myInvMass {pipi}}
             {\ifthenelse{\equal{#1}{Bd2KSpiK} \or \equal{#1}{Bs2KSpiK}}    {\def\myInvMass {piK}}
               {\ifthenelse{\equal{#1}{Bd2KSKpi} \or \equal{#1}{Bs2KSKpi}}  {\def\myInvMass {Kpi}}
                 {\ifthenelse{\equal{#1}{Bd2KSKK} \or \equal{#1}{Bs2KSKK}}  {\def\myInvMass {KK}}{\def\myInvMass ERROR}}
               }
             }

}
\newcommand{\makeMode}[1]{
  \ifthenelse{\equal{#1}{Bd2KSpipi}}                     {\def\myMode {\BdtoKsPiPi}}
             {\ifthenelse{\equal{#1}{Bd2KSpiK}}          {\def\myMode {\BdtoKsPipKm}}
               {\ifthenelse{\equal{#1}{Bd2KSKpi}}        {\def\myMode {\BdtoKsKpPim}}
                 {\ifthenelse{\equal{#1}{Bd2KSKK}}       {\def\myMode {\BdtoKsKK}}
                   {\ifthenelse{\equal{#1}{Bs2KSpipi}}           {\def\myMode {\BstoKsPiPi}}
                     {\ifthenelse{\equal{#1}{Bs2KSpiK}}          {\def\myMode {\BstoKsPipKm}}
                       {\ifthenelse{\equal{#1}{Bs2KSKpi}}        {\def\myMode {\BstoKsKpPim}}
                         {\ifthenelse{\equal{#1}{Bs2KSKK}}       {\def\myMode {\BstoKsKK}}{\def\myMode{ERROR}}
                         }
                       }
                     }
                   }
                 }
               }
             }
}

\newcommand{\plotGeomEff}[3]{
  \makeInvMass{#2}
  \makeMode{#2}
  \ifthenelse{\equal{#1}{2011}}{\def\redYear {2011}}{\def\redYear {2012}}
  \centering
  \includegraphics[width=0.45\textwidth]{figs/EfficiencyMaps/#1/Geometry/NoSel/#2/Efficiency_Map_NoSel_#2_#3_#1_\myInvMass.pdf.eps}
  \includegraphics[width=0.45\textwidth]{figs/EfficiencyMaps/#1/Geometry/NoSel/#2/Spline_Eff_NoSel_#2_#3_#1_\myInvMass.pdf.eps}
  \includegraphics[width=0.45\textwidth]{figs/EfficiencyMaps/#1/Geometry/NoSel/#2/Efficiency_errorHi_Map_NoSel_#2_#3_#1_\myInvMass.pdf.eps}
  \includegraphics[width=0.45\textwidth]{figs/EfficiencyMaps/#1/Geometry/NoSel/#2/Efficiency_errorLo_Map_NoSel_#2_#3_#1_\myInvMass.pdf.eps}
  \caption{
    (Top) $\epsilon^{\rm geom}$ as a function of the \myMode
    square Dalitz plot position obtained from \redYear-conditions generator-level signal MC:
    (left) the raw histogram,
    (right) smoothed using a 2D cubic spline.
    (Bottom) the (left) upper and (right) lower uncertainties on the histogram bins.
    Uncertainties are due to MC statistics.
  }
  \label{fig:geoeff:GeoEff:#1:#2:#3}
}

\newcommand{\plotTrackCorr}[4]{
  \makeInvMass{#2}
  \makeMode{#2}
  \centering
  \includegraphics[width=0.45\textwidth]{figs/EfficiencyMaps/#1/Tracking/#4/#2/Trk_AllEff-#4-#2-#3-#1-\myInvMass.pdf.eps}
  \includegraphics[width=0.45\textwidth]{figs/EfficiencyMaps/#1/Tracking/#4/#2/Spline_Eff_#4_#2_#3_#1_\myInvMass.pdf.eps}\\
  \includegraphics[width=0.45\textwidth]{figs/EfficiencyMaps/#1/Tracking/#4/#2/Tracking_errorHi_bootstrap_#4_#2_#3_\myInvMass_#1.pdf.eps}
  \includegraphics[width=0.45\textwidth]{figs/EfficiencyMaps/#1/Tracking/#4/#2/Tracking_errorHi_bootstrap_#4_#2_#3_\myInvMass_#1.pdf.eps}
  \caption{
    (Top) Combined tracking efficiency corrections in the \myMode signal mode for #3 and #1 configuration:
    (left) the raw histogram obtained from MC simulation and (right) smoothed using a 2D cubic spline.  
    (Bottom) the (left) upper and (right) lower uncertainties on the histogram bins.
  }
  \label{fig:DPefficiency-trk-all-#1-#2-#3-#4}
}

\newcommand{\plotTrackCorrMode}[1]{
  \begin{figure}[!htb]
    \plotTrackCorr{2011}{#1}{DD}{Loose}
  \end{figure}
  \begin{figure}[!htb]
    \plotTrackCorr{2011}{#1}{LL}{Loose}
  \end{figure}
}

\newcommand{\plotTrigCorr}[4]{
  \makeInvMass{#1}
  \makeMode{#1}
  \ifthenelse{\equal{#4}{TOS}}{
    \def\trigComment {$\epsilon^{\rm L0TOS|sel\&geom}_{\rm data}/\epsilon^{\rm L0TOS|sel\&geom}_{\rm MC}$}}{
    \def\trigComment {$\epsilon^{\rm !L0TOS|sel\&geom}_{\rm data}/\epsilon^{\rm !L0TOS|sel\&geom}_{\rm MC}$}}
  
  \centering
  \includegraphics[width=0.45\textwidth]{figs/EfficiencyMaps/2011/L0#4/#3/#1/L0#4-Correction-#3-#1-#2-2011-\myInvMass.pdf.eps}
  \includegraphics[width=0.45\textwidth]{figs/EfficiencyMaps/2011/L0#4/#3/#1/Spline_Eff_#3_#1_#2_2011_\myInvMass.pdf.eps}\\
  \includegraphics[width=0.45\textwidth]{figs/EfficiencyMaps/2011/L0#4/#3/#1/L0#4_errorHi_combined_#3_#1_#2_\myInvMass_2011.pdf.eps}
  \includegraphics[width=0.45\textwidth]{figs/EfficiencyMaps/2011/L0#4/#3/#1/L0#4_errorLo_combined_#3_#1_#2_\myInvMass_2011.pdf.eps}
  \caption{
    (Top) \trigComment across the \myMode #2 square Dalitz plot (2011+2012 combined):
    (left) the raw histogram obtained using the procedure described in the text and (right) smoothed using a 2D cubic spline.  
    (Bottom) the (left) upper and (right) lower uncertainties on the histogram bins.
  }
  \label{fig:DPcorrection-#1-#2-#3-#4}
}

\newcommand{\plotSelEff}[5]{
  \makeInvMass{#2}
  \makeMode{#2}
  \ifthenelse{\equal{#5}{TOS}}{\def\selComment {{\tt L0Hadron\_TOS}}}{\def\selComment {{\tt L0Global\_TIS\&\&!L0Hadron\_TOS}}}
  \centering
  \includegraphics[width=0.45\textwidth]{figs/EfficiencyMaps/#1/Sel#5/#4/#2/#2-Eff-#4-#2-#3-#1-\myInvMass.pdf.eps}
  \includegraphics[width=0.45\textwidth]{figs/EfficiencyMaps/#1/Sel#5/#4/#2/Spline_Eff_#4_#2_#3_#1_\myInvMass.pdf.eps}\\
  \includegraphics[width=0.45\textwidth]{figs/EfficiencyMaps/#1/Sel#5/#4/#2/#2-ErrorHi-#4-#2-#3-#1-\myInvMass.pdf.eps}
  \includegraphics[width=0.45\textwidth]{figs/EfficiencyMaps/#1/Sel#5/#4/#2/#2-ErrorLo-#4-#2-#3-#1-\myInvMass.pdf.eps}
  \caption{
    (Top) $\epsilon^{\rm sel|geom}$ across the \myMode #3 #1 square Dalitz plot for \selComment \MakeLowercase{#4} BDT candidates:
    (left) the raw histogram obtained using the procedure described in the text and (right) smoothed using a 2D cubic spline.  
    (Bottom) the (left) upper and (right) lower uncertainties on the histogram bins.
  }
  \label{fig:DPefficiency-sel-#1-#2-#3-#4-#5}
}

\newcommand{\plotSelEffMode}[2]{
  \begin{figure}[!htb]
    \plotSelEff{2011}{#1}{DD}{#1}{TOS}
  \end{figure}
  \begin{figure}[!htb]
    \plotSelEff{2011}{#1}{DD}{#1}{TIS}
  \end{figure}
  \begin{figure}[!htb]
    \plotSelEff{2011}{#1}{LL}{#1}{TOS}
  \end{figure}
  \begin{figure}[!htb]
    \plotSelEff{2011}{#1}{LL}{#1}{TIS}
  \end{figure}
}

\newcommand{\fullDataTaking}[1]
           {
             \ifthenelse{\equal{#1}{2011}}
                        {\def\fullYear {2011\xspace}}
                        {\ifthenelse{\equal{#1}{2012a}}{\def\fullYear {2012a\xspace}}{\def\fullYear {2012b\xspace}}}
           }
\newcommand{\selOptChoice}[1]
           {
	     \ifthenelse{\equal{#1}{Loose}}
	                {\def\selOpt {favoured\xspace}}
			{\def\selOpt {suppressed\xspace}}
	   }
\newcommand{\paperFitResults}[2]
           {
             \selOptChoice{#1}
               \ifthenelse{\equal{#1}{Loose}}{
               \def\myCaption{In each plot, data are the black points with error bars and the total fit model is overlaid (solid blue line).
                 The \Bd (\Bs) signal components are the pink (orange) short-dashed (dotted) lines,
                 while fully reconstructed misidentified decays are the green and dark blue dashed lines
                 close to the \Bd and \Bs peaks.
                 The sum of the partially reconstructed contributions from \B to open charm decays,
                 charmless hadronic decays, \BdtoetapKs and charmless radiative decays are
                 the red dash triple-dotted lines.
                 The combinatorial background contribution is the gray long-dash dotted
                 line.}
                 }{
                 \def\myCaption{Colours and line styles follow the same conventions as in Fig.~\ref{fig : fitLoose2011}} }
             \fullDataTaking{#2}
             \begin{figure}[!htbp]
               \begin{center}
                 \ifthenelse{\boolean{pdflatex}}{

                   \includegraphics*[width=0.49\textwidth]{figs/FusedCats-#1-KSKKDD-log.pdf}
                   \includegraphics*[width=0.49\textwidth]{figs/FusedCats-#1-KSKKLL-log.pdf}\\

                   \includegraphics*[width=0.49\textwidth]{figs/FusedCats-#1-KSKpiDD-log.pdf}
                   \includegraphics*[width=0.49\textwidth]{figs/FusedCats-#1-KSKpiLL-log.pdf}\\
                  
                   \includegraphics*[width=0.49\textwidth]{figs/FusedCats-#1-KSpipiDD-log.pdf}
                   \includegraphics*[width=0.49\textwidth]{figs/FusedCats-#1-KSpipiLL-log.pdf}

                 }{

                   \includegraphics*[width=0.49\textwidth]{figs/FusedCats-#1-KSKKDD-log.eps}
                   \includegraphics*[width=0.49\textwidth]{figs/FusedCats-#1-KSKKLL-log.eps}\\

                   \includegraphics*[width=0.49\textwidth]{figs/FusedCats-#1-KSKpiDD-log.eps}
                   \includegraphics*[width=0.49\textwidth]{figs/FusedCats-#1-KSKpiLL-log.eps}\\

                   \includegraphics*[width=0.49\textwidth]{figs/FusedCats-#1-KSpipiDD-log.eps}
                   \includegraphics*[width=0.49\textwidth]{figs/FusedCats-#1-KSpipiLL-log.eps}

                 }
               \end{center}
                 \caption{
                 Invariant mass distributions of, from top to bottom, \KsKK, \KsKPi
                 and \KsPiPi candidates, summing the three periods of
                 data taking, with the selection optimisation chosen
                 for the \selOpt decay modes for (left) \DD~and
                 (right) \LL\ \KS reconstruction categories.
                 \myCaption
                  }                

               \label{fig : fit#1#2}
             \end{figure}
           }

\newcommand\paperCaption[1]{
  \fullDataTaking{#1}
  \caption{Signal yields obtained for the \fullYear category from the
    simultaneous fit to the data.
    The yields shown are those obtained when fitting the data sample
    selected using the BDT optimisation chosen for the given decay mode.
    The uncertainties are statistical only.
    The average selection efficiencies, described in
    Sec.~\ref{sec:Efficiencies}, are also shown for each decay mode
    together with the corresponding total uncertainty due to the limited
    simulation sample size and systematic effects in their determination.
  }
  \label{tab : FitResults#1}

}

%% file: title-LHCb-PAPER.tex

\begin{titlepage}
\pagenumbering{roman}

\vspace*{-1.5cm}
\centerline{\large EUROPEAN ORGANIZATION FOR NUCLEAR RESEARCH (CERN)}
\vspace*{1.5cm}
\noindent
\begin{tabular*}{\linewidth}{lc@{\extracolsep{\fill}}r@{\extracolsep{0pt}}}
\ifthenelse{\boolean{pdflatex}}
{\vspace*{-2.7cm}\mbox{\!\!\!\includegraphics[width=.14\textwidth]{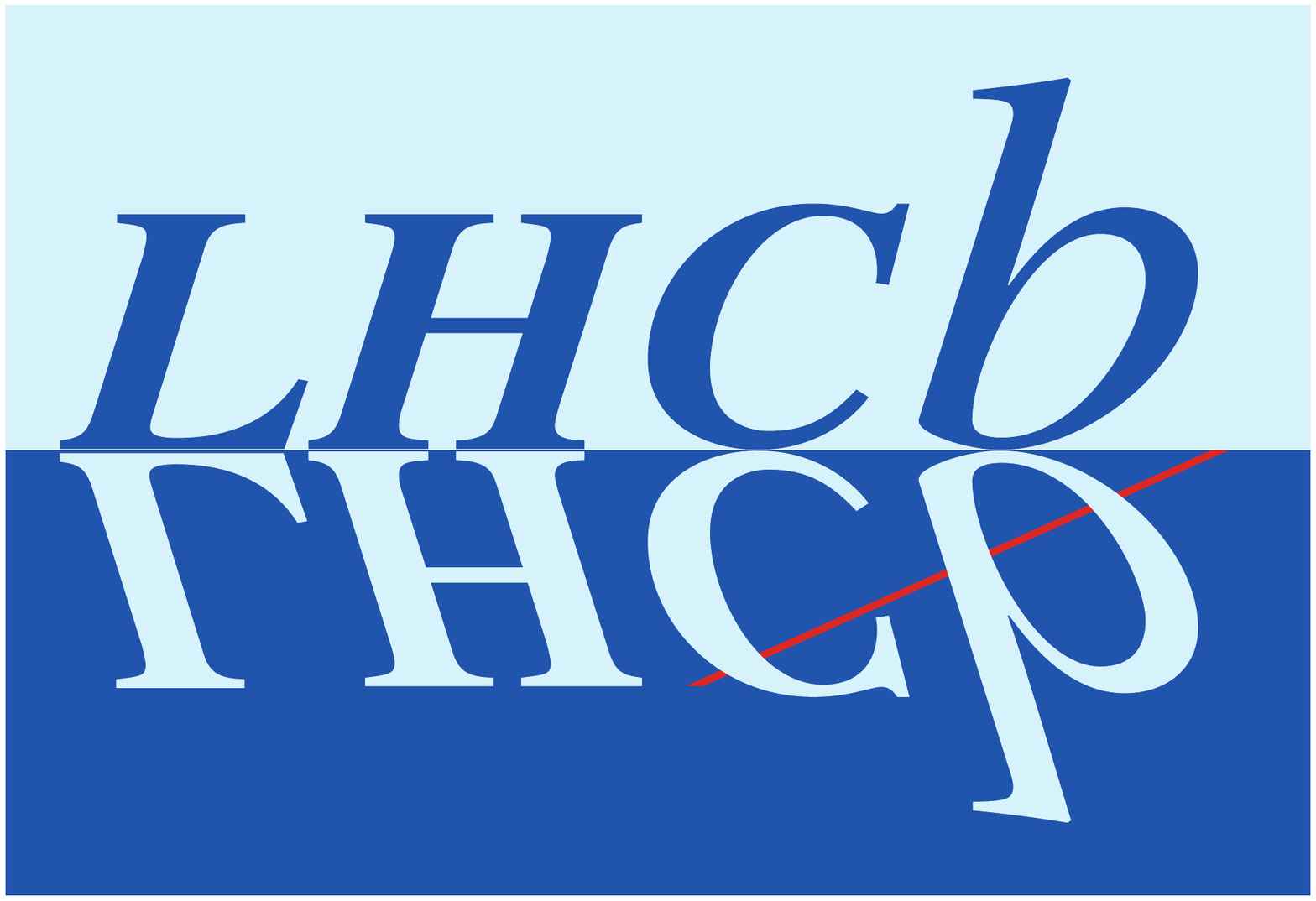}} & &}%
{\vspace*{-1.2cm}\mbox{\!\!\!\includegraphics[width=.12\textwidth]{lhcb-logo.eps}} & &}%
\\
 & & CERN-EP-2017-131 \\  
 & & LHCb-PAPER-2017-010 \\  
 & & 05 July 2017 \\ 
 & & \\
\end{tabular*}

\vspace*{1.5cm}

{\normalfont\bfseries\boldmath\LARGE
\begin{center}
Updated branching fraction measurements of $B^0_{(s)} \to K_{\mathrm{\scriptscriptstyle S}}^0 h^+ h^{\prime -}$ decays
\end{center}
}

\vspace*{1.0cm}

\begin{center}
The LHCb collaboration\footnote{Authors are listed at the end of this article.}
\end{center}

\vspace{0.5\fill}

\begin{abstract}
  \noindent
  The charmless three-body decays $B^0_{(s)} \to K_{\mathrm{\scriptscriptstyle S}}^0 h^+ h^{\prime -}$
  (where $h^{(\prime)} = \pi, K$) are analysed using a sample of $pp$ collision data recorded by the LHCb
  experiment, corresponding to an integrated luminosity of $3\mbox{\,fb}^{-1}$.
  The branching fractions are measured relative to that of the
  $B^0 \to K_{\mathrm{\scriptscriptstyle S}}^0 \pi^{+} \pi^{-}$ decay,
  and are determined to be:
  \begin{align*}
  \frac{\Br{\BdtoKsKPi}} {\Br{\BdtoKsPiPi}}  = {} & 0.123 \pm 0.009 \; \stat \; \pm 0.015 \; \syst \,,\\ 
  \frac{\Br{\BdtoKsKK}}  {\Br{\BdtoKsPiPi}}  = {} & 0.549 \pm 0.018 \; \stat \; \pm 0.033 \; \syst \,,\\
  \frac{\Br{\BstoKsPiPi}}{\Br{\BdtoKsPiPi}}  = {} & 0.191 \pm 0.027 \; \stat \; \pm 0.031 \; \syst \; \pm 0.011 \; (f_s/f_d) \,, \\
  \frac{\Br{\BstoKsKPi}} {\Br{\BdtoKsPiPi}}  = {} & 1.70\phantom{0} \pm 0.07\phantom{0} \; \stat \; \pm  0.11\phantom{0} \; \syst \; \pm 0.10\phantom{0} \; (f_s/f_d) \,,\\
  \frac{\Br{\BstoKsKK}}{\Br{\BdtoKsPiPi}}  \in {} & [0.008 - 0.051] \mathrm{~at~90\%~confidence~level,}
  \end{align*}  
  where $f_s/f_d$ represents the ratio of hadronisation fractions of the $B^0_s$ and $B^0$ mesons.
\end{abstract}
\vspace*{1.0cm}

\begin{center}
  Published as J. High Energ. Phys. (2017) 2017: 27
\end{center}

\vspace{\fill}

{\footnotesize 
\centerline{\copyright~CERN on behalf of the \lhcb collaboration, licence \href{http://creativecommons.org/licenses/by/4.0/}{CC-BY-4.0}.}}
\vspace*{2mm}

\end{titlepage}


\newpage
\setcounter{page}{2}
\mbox{~}
%
%
%
%

\cleardoublepage

%% file: introduction.tex
\section{Introduction}
\label{sec:Introduction}

The measurement of \CP-violation observables in the decays \BdtoKsPiPi and
\mbox{\BdtoKsKK},
which are dominated by \btoqqbars ($q = u,d,s$) loop transitions, are of great
theoretical interest.\footnote{Unless stated otherwise, charge
conjugated modes are implicitly included throughout this article.}
In particular, the mixing-induced \CP asymmetries in these decays are
predicted by the Standard Model (SM) Cabibbo--Kobayashi--Maskawa
mechanism~\cite{Cabibbo:1963yz,Kobayashi:1973fv} to be approximately equal
to those governed by \btoccbars transitions, such as $\Bd\to\jpsi\KS$.
Within the SM the weak phase measurements in
\btoqqbars decays are expected to deviate from the values determined in
\btoccbars decays but for certain contributions to these decays, such as
$\Bd\to\phi\KS$ and $\Bd\to\rhoz\KS$, this deviation is either expected to
be small or can be controlled using flavour
symmetries~\cite{Beneke:2005pu,Buchalla:2005us,Dutta:2008xw}.
The existence of new particles predicted in several extensions of the
SM could introduce additional weak phases that contribute along with the SM
mixing phase to the amplitudes of these loop-dominated charmless
decays, potentially leading to much greater deviations from the \btoccbars
values~\cite{Grossman:1996ke,London:1997zk,Ciuchini:1997zp}.
The mixing-induced \CP-violating phase can be measured by means of a
flavour-tagged time-dependent analysis of the three-body Dalitz plot of these
decays~\cite{Dalseno:2008wwa,Aubert:2009me,Nakahama:2010nj,Lees:2012kxa}.   
The current experimental measurements of this phase in \btoqqbars decays~\cite{HFAG} show
a generally good agreement with the results for the weak phase \Pbeta from \btoccbars decays
for each of the \CP eigenstates studied.
The experimental uncertainties are, however, currently rather larger than
the size of the expected deviations, both in the SM and beyond-the-SM
scenarios, and so there is a need for more precise measurements of
these quantities.
A similar determination of the mixing-induced \CP-violating phase in the \Bs
system is possible with, among others, the \BstoKsKPi
decays~\cite{Ciuchini:2007hx}.

It is also possible to determine the CKM angle \Pgamma by combining
information from several $\B \to \kaon\had\hadprim$ decays, using either the
methods originally proposed in Refs.~\cite{Ciuchini:2006kv,Gronau:2006qn}
and recently developed further in Ref.~\cite{Charles:2017ptc}, or those
proposed in Refs.~\cite{ReyLeLorier:2011ww,Bhattacharya:2013cla,Bhattacharya:2015uua}.
The existing experimental results, which come from the
\babar collaboration~\cite{BABAR:2011ae,Lees:2015uun}, demonstrate the feasibility of the
measurement, albeit with large statistical uncertainties.
The decay \BstoKsPiPi is dominated by tree-level processes and as such is of
particular interest for this effort, with the potential to yield a
theoretically clean determination of \Pgamma~\cite{Ciuchini:2006st}.  

The measurements of the branching fractions themselves are of great
importance in order to confront theoretical predictions.
These predictions are based on various approaches to modelling the
hadronisation processes, such as QCD factorisation or PQCD, see for example
Refs.~\cite{Cheng:2013dua,Cheng:2014uga,Li:2014fla,Li:2014oca,Wang:2016rlo,Li:2017mao}.
Comparison of the different approaches with the experimental data will
allow further refinement of the theoretical models, which in turn will
yield improved predictions of branching fractions and \CP asymmetries of
these and many other charmless decay modes.
In addition, these results can be used to test the level of breaking of the flavour
symmetries: isospin, U-spin and SU(3), see for example Ref.~\cite{He:2014xha}.

Of the decays of neutral \B mesons to \KsPiPi, \KsKPi and \KsKK final
states, only the decay \BstoKsKK remains to be observed~\cite{Garmash:2003er,Garmash:2006fh,Aubert:2009me,delAmoSanchez:2010ur,Lees:2012kxa,LHCb-PAPER-2013-042}. 
Most recently, a search for the three \Bs decays was reported by the \lhcb
experiment using the 1\invfb data sample recorded in
2011~\cite{LHCb-PAPER-2013-042}.
While first observations were made for the \BstoKsPiPi and \BstoKsKPi modes,
no evidence for the decay \BstoKsKK was found.
In this work, all the aforementioned charmless three-body decays
of the \Bd and \Bs mesons are studied using the $pp$ collision data
recorded by the LHCb detector, corresponding to an integrated luminosity of
1.0\invfb at a centre-of-mass energy of 7\tev in 2011 and 2.0\invfb at a
centre-of-mass energy of 8\tev in 2012.
This sample is three times larger than that used in Ref.~\cite{LHCb-PAPER-2013-042}. 
The measurements of the time-integrated branching fractions~\cite{DeBruyn:2012wj} 
relative to that of \BdtoKsPiPi are presented. 
The notation \Br{\Bz\to\KS\Kpm\pimp} is used throughout the document to
indicate the sum of the branching fractions \Br{\Bz\to\KS\Kp\pim} and
\Br{\Bz\to\KS\Km\pip}, and similarly for the corresponding \Bs decays.

%% file: detector.tex
\section{Detector and simulation}
\label{sec:Detector}

The \lhcb detector~\cite{Alves:2008zz,LHCb-DP-2014-002} is a single-arm
forward spectrometer covering the \mbox{pseudorapidity} range $2<\eta <5$,
designed for the study of particles containing \bquark or \cquark quarks.
The detector includes a high-precision tracking system consisting of a
silicon-strip vertex detector (\velo) surrounding the $pp$ interaction
region,
a large-area silicon-strip detector located
upstream of a dipole magnet with a bending power of about
$4{\mathrm{\,Tm}}$, and three stations of silicon-strip detectors and straw
drift tubes
placed downstream of the magnet.
The tracking system provides a measurement of momentum, \ptot, of charged particles with
a relative uncertainty that varies from 0.5\% at low momentum to 1.0\% at 200\gevc.
The minimum distance of a track to a primary vertex (PV), the impact parameter (IP), 
is measured with a resolution of $(15+29/\pt)\mum$,
where \pt is the component of the momentum transverse to the beam, in\,\gevc.
Different types of charged hadrons are distinguished using information
from two ring-imaging Cherenkov detectors.
Photons, electrons and hadrons are identified by a calorimeter system consisting of
scintillating-pad and preshower detectors, an electromagnetic
calorimeter and a hadronic calorimeter. Muons are identified by a
system composed of alternating layers of iron and multiwire
proportional chambers.

Simulated data samples are used to investigate backgrounds from other
\bquark-hadron decays and also to study the detection and reconstruction
efficiency of the signal.
In the simulation, $pp$ collisions are generated using
\pythia~\cite{Sjostrand:2006za,*Sjostrand:2007gs} with a specific \lhcb
configuration~\cite{LHCb-PROC-2010-056}.
Decays of hadronic particles are described by \evtgen~\cite{Lange:2001uf},
in which final-state radiation is generated using
\photos~\cite{Golonka:2005pn}.
The interaction of the generated particles with the detector, and its
response, are implemented using the \geant toolkit~\cite{Allison:2006ve,
*Agostinelli:2002hh} as described in Ref.~\cite{LHCb-PROC-2011-006}.

%% file: selection.tex
\section{Trigger and event selection}
\label{sec:Selection}

The online event selection is performed by a
trigger~\cite{LHCb-DP-2012-004}, 
which consists of a hardware stage, based on information from the calorimeter and muon
systems, followed by a software stage, in which all charged particles
with $\pt>500\,(300)\mevc$ are reconstructed for data collected in 2011\,(2012).
At the hardware trigger stage, events are required to have a muon with high
\pt or a hadron, photon or electron with high transverse energy in the
calorimeters.
The software trigger requires a two-, three- or four-track secondary vertex
with a significant displacement from all primary $pp$ interaction vertices.
At least one charged particle must have transverse momentum $\pt >
1.7\,(1.6)\gevc$ in the 2011\,(2012) data and be inconsistent with
originating from a PV.
A multivariate algorithm~\cite{BBDT} is used for the identification of
secondary vertices consistent with the decay of a \bquark hadron.
It is required that the software trigger decision must have been caused
entirely by tracks from the decay of the signal \B candidate.

To suppress the `combinatorial' background formed by combinations of unrelated
tracks, the events satisfying the trigger requirements are filtered in two
stages: a preselection based on loose requirements, followed by a multivariate selection.
In order to minimise the variation of the selection efficiency over the
Dalitz plot, the selection procedure uses only loose requirements on the
momenta of the \B-meson decay products and relies mainly on topological
features such as the flight distance of the \B candidate. 
These features depend on whether the \B candidate or the \KS, \hadp,
\hadprimm candidates are consistent with having originated from a
particular PV.
It is therefore necessary to `associate' each candidate with a single PV
--- that from which it is most consistent with having originated.
The association is defined in terms of the \chisqip quantity, which is the
difference in fit \chisq of the given PV reconstructed with and without the
track or tracks from the particle in question.
In events that contain more than one PV, each candidate is associated with
the PV that has the smallest \chisqip.

Decays of \decay{\KS}{\pip\pim} are reconstructed in two different categories:
the first involving \KS mesons that decay early enough for the
resulting pions to be reconstructed in the \velo; and the
second containing those \KS mesons that decay later, such that track
segments of the pions cannot be formed in the \velo.
These \KS reconstruction categories are referred to as \emph{\LL} and \emph{\DD}, respectively.
The \LL\ category has better mass, momentum and vertex resolution than the
\DD\ category.
There are however approximately twice as many \KS candidates reconstructed
as \DD\ than as \LL, simply due to the lifetime of the \KS meson and the
geometry of the detector.
In the following, \B candidates reconstructed from either a \LL\ or \DD\
\KS candidate, in addition to two oppositely charged tracks, are also
referred to using these category names.
During the 2012 data taking, a significant improvement of the trigger
efficiency for long-lived particles, specifically for downstream
candidates, was obtained following an update of the software trigger
algorithms.
To take into account the differences in trigger efficiencies and the
different data-taking conditions, the data sample is divided into 2011,
2012a, and 2012b data-taking periods, and each period is divided in two
sub-samples according to the \KS reconstruction category.
The 2012b sample is the largest, corresponding to 1.4\invfb, and also has
the highest trigger efficiency.

The two charged pions that form the \KS candidates are both required to
have momentum $\ptot>2\gevc$ and have \chisqip with respect to their
associated PV greater than 9 (4) for \LL\ (\DD) candidates.
They are then required to form a vertex with good fit quality (quantified
by the fit \chisq, $\chisqvtx < 12$)
and to have invariant mass within 20\mevcc (30\mevcc) of the nominal \KS
mass~\cite{PDG2016} for \LL\ (\DD) candidates.
A requirement on the square of the ratio of the separation of the \KS
vertex from its associated PV and the corresponding uncertainty,
$\chisqvs > 80 \, (50)$ for \LL\ (\DD) candidates, ensures a significant
vertex separation.
Downstream \KS candidates are required in addition to have a momentum $\ptot>6\gevc$.

The \B candidates are formed from a \KS candidate and two oppositely
charged tracks (initially reconstructed under the pion mass hypothesis).
Each of these two tracks is required to have $\ptot<100\gevc$, a
value beyond which there is little pion-kaon discrimination.
The scalar sum of the transverse momenta of the \KS and the two
$\hadp\hadprimm$ candidates must be greater than 3.0\gevc (4.2\gevc), for \LL\ (\DD)
candidates, and at least two of the three decay products must have $\pt>0.8\gevc$.
The IP of the \B-meson decay product with the largest \pt is required to be
greater than 0.05\mm relative to the PV associated to the \B candidate.
The \B candidate decay products are then required to form a vertex that has
$\chisqvtx<12$ and which is separated from any PV by at least 1.7\mm.
The difference in \chisqvtx when adding another track must be greater than 4.
The \B candidates must have $\pt>1.5\gevc$ and invariant mass within
the range $4000 < m_{\KsPiPi} < 6200 \mevcc$.
They are further obliged to be consistent with originating from a PV,
quantified by requiring, for \LL\ (\DD) candidates, both that
$\chisqip < 8 \, (6)$ and that the cosine of the angle $\theta_{\rm DIR}$ between
the reconstructed momentum of the \B candidate and the vector between the
associated PV and the decay vertex be greater than 0.9999 (0.999).
Finally, the decay vertex of the \KS candidate is required to be at least
30\mm downstream, along the beam direction, from that of the \B candidate.

Multivariate discriminants based on a boosted decision tree~(BDT)
algorithm~\cite{Breiman,AdaBoost}
are used to further reduce combinatorial backgrounds.
Simulated \BdtoKsPiPi events and data from upper mass sidebands,
$5425 < m_{\KsPiPi} < 6200 \mevcc$,
are used as the signal and background training samples, respectively.
Contributions from muons and protons are removed from these samples
using particle identification (PID) variables.
Each of the six samples (resulting from the
division by the three data-taking periods and the two 
\KS reconstruction categories) is further subdivided into two equally-sized
subsamples.  Each subsample is then used to train an independent
discriminant.
In the subsequent analysis the BDT trained on one subsample of a given
category is used to select events from the other subsample, in order to
avoid bias.
The input quantities for the BDTs are:
the \pt, $\eta$, \chisqip, \chisqvs, $\cos\theta_{\rm DIR}$ and \chisqvtx values of the \B candidate;
the smallest change in the \B-candidate \chisqvtx value when adding another track from the event;
the sum of the \chisqip values of the \hadp and \hadm candidates;
the \chisqip, \chisqvs and \chisqvtx values of the \KS candidate;
and the \pt asymmetry
\begin{equation}
\pt^\mathrm{asym} \equiv \frac{\pt^\B - \pt^\mathrm{cone}}{\pt^\B + \pt^\mathrm{cone}} \,,
\end{equation}
where $\pt^\mathrm{cone}$ is the transverse component of the sum of all
particle momenta inside a cone around the \B-candidate direction, of
radius $R \equiv \sqrt{\delta\eta^2 + \delta\phi^2} = 1.5$,
where $\delta\eta$ and $\delta\phi$ are the difference in pseudorapidity
and azimuthal angle (in radians) around the beam direction,
between the momentum vector of the track under consideration and that of the \B candidate.

The selection requirement placed on the output of the BDTs is independently
optimised for each data sample. 
For all signal decay modes that have previously been observed, the following
figure of merit is used 
\begin{equation}
{\cal Q}_1 \equiv \frac{N_{\rm sig}}{\sqrt{N_{\rm sig}+N_{\rm bg}}} \,,
\end{equation}
where $N_{\rm sig}$ ($N_{\rm bg}$) represents the number of expected signal
(combinatorial background) events for a given selection.
The value of $N_{\rm sig}$ is estimated based on the known branching fractions
and efficiencies, while $N_{\rm bg}$ is calculated by fitting the sideband above
the signal region and extrapolating into the signal region, defined as the
invariant-mass window of five times the typical resolution around the \Bd
and the \Bs masses.

For the yet unobserved  \BstoKsKK mode, an alternative figure
of merit~\cite{Punzi:2003bu} is used
\begin{equation}
{\cal Q}_2 \equiv \frac{\eps_{\rm sig}}{1 + \sqrt{N_{\rm bg}}} \,,
\end{equation}
where the signal efficiency ($\eps_{\rm sig}$) is estimated from
the signal simulation.
The optimisation is performed separately for each of the six categories.
As each final state contains both \Bd and \Bs signals, one of which is favoured and
the other suppressed,
this procedure results in applying two differently optimised selections on each final state.

Particle identification requirements are subsequently applied in order to reduce
backgrounds from decays such as \LbtoKsPip and
\decay{\Bdsz}{\subdecay{\jpsi}{\mup\mun}\KS} where, respectively, the proton and
muons are misidentified as pions or kaons.
PID information is also used to assign each candidate exclusively to one out 
of four possible final states: \KsPiPi, \KsKpPim, \KsPipKm, and \KsKK.
The PID requirements are optimised to reduce the cross-feed between the different signal
decay modes using the same figures of merit
introduced for the BDT optimisation. 

Fully reconstructed \B-meson decays into two-body $\D\had$ or
$(\cquark\cquarkbar)\KS$ combinations, where $(\cquark\cquarkbar)$
indicates a charmonium resonance, may result in a \Kshhp final state that
satisfies the selection criteria and has the same \B-candidate
invariant mass distribution as the signal candidates.
The decays of \Lb baryons to $\Lc\had$ with \decay{\Lc}{\proton\KS} also
peak under the signal when the proton is misidentified.
Therefore, the following \D and \Lc decays are explicitly reconstructed
under the relevant particle hypotheses and vetoed in all the spectra:
${\Dz \to \Km\pip}$,
${\Dp \to \KS\Kp}$,
${\Dp \to \KS\pip}$,
${\Dsp \to \KS\Kp}$,
${\Dsp \to \KS\pip}$,
and
${\Lc \to \proton\KS}$.
Additional vetoes on charmonium resonances, ${\jpsi \to \pipi, \Kp\Km}$ and
${\chiczero \to \pipi, \Kp\Km}$, are applied to remove the small number of fully
reconstructed and well identified peaking
${\decay{\Bdsz}{\left(\jpsi,\chiczero\right)\KS}}$ decays.
The vetoed region for each reconstructed charm (charmonium) state is an
invariant-mass window of $30\,(48)\mevcc$ around the world average mass
value of that state~\cite{PDG2016}.
This range reflects the typical mass resolution obtained at \lhcb.

The fraction of selected events containing more than one \B candidate is at
the percent level.  The candidate to be retained in each event is chosen
randomly, but reproducibly.

%% file: fit-model.tex
\section{Fit model}
\label{sec:FitModel}

The signal yields corresponding to each of the BDT
optimisations are determined by means of a simultaneous unbinned extended
maximum likelihood fit to the \B-candidate invariant mass distributions of
all final states in the six categories.
Four types of components contribute to each invariant mass distribution: signal
decays, backgrounds resulting from cross-feeds, partially reconstructed decays,
and random combinations of unrelated tracks.

Signal \BdstoKshhp decays with correct identification of the final-state
particles are modelled with the sum of two Crystal Ball (CB)
functions~\cite{Skwarnicki:1986xj} that share common values for the peak
position and width but have independent power law tails on opposite sides
of the peak.
The \Bd and \Bs masses (peak positions of the CB functions) are free
parameters in the fit and are allowed to take different values in the
different data-taking periods in order to allow for small differences in
momentum calibration.
Seven parameters related to the widths of the CB functions
are also free parameters of the fit:
the width of the \DD\ \BdtoKsPiPi signal in each of the three data-taking periods;
the ratio of the widths of the \Bs and \Bd decay modes;
the relative widths of \KsKPi and \KsKK to \KsPiPi;
and the ratio of the widths in the \LL\ and \DD\ categories.
The dependence of the width on each of these divisions is assumed to
factorise; for example the width $\sigma$ of the \LL\ \BstoKsKK signal in the 2011
data-taking period is related to that of the downstream \BdtoKsPiPi signal in the same data-taking period by
\begin{equation}
\sigma^{\rm 2011\,\LL}_{\BstoKsKK} = \sigma^{\rm 2011\,\DD}_{\BdtoKsPiPi}
\times r_{\Bs/\Bd} \times r_{\KsKK/\KsPiPi} \times r_{\rm \LL/\DD} \, ,
\end{equation}
where $r_{x/y}$ indicates the ratio of the widths of categories $x$ and $y$.
These assumptions are made necessary by the otherwise poor determination of
the width of the suppressed mode in each spectrum.
The other parameters of the CB components are obtained by a simultaneous
fit to simulated samples.

Cross-feed contributions from misidentified signal decays are
modelled empirically by the sum of two CB functions using simulated events.
Only contributions from the decays \BdtoKsPiPi and \BdtoKsKK reconstructed
and selected as \KsKPi, or the decays \BstoKsKPi and \BdtoKsKPi
reconstructed and selected as either \KsKK or \KsPiPi are considered.
Other potential misidentified decays are neglected, as their contributions have been
checked to be below one event.
The relative yield of each misidentified decay is constrained with respect
to the yield of the corresponding correctly identified decay.
The constraints are implemented using Gaussian prior probability
distributions included in the
likelihood. The mean values are obtained from the ratio of selection
efficiencies and the widths include uncertainties originating from 
the finite size of the simulated event samples and the systematic 
uncertainties related to the determination of the PID efficiencies.

Backgrounds from partially reconstructed decays such as \mbox{\BstoKstzKstzbtoKsPizKPi},
where the neutral pion is not reconstructed, are also modelled.
Four categories are included in each of the final state spectra, where the
background results from either charmed or charmless decays of $\B^{0,+}$ or
\Bs mesons.
These decays are modelled by means of generalised ARGUS
functions~\cite{Albrecht:1990cs} convolved with a Gaussian resolution
function.
Their parameters are determined from simulated samples of the expected dominant
decays in each category.
Radiative decays and those from \BdtoetapKs are considered separately and
included only in the \KsPiPi final state.
The normalisation of all such contributions is constrained with respect to
the signal in the relevant final state using Gaussian prior probability
distributions based on the ratio of efficiencies and the ratio of branching
fractions from world averages~\cite{PDG2016}. The relative uncertainties on
 these ratios vary between 20\% and 100\%.

The combinatorial background is modelled by a linear function.
The variations of the slope parameter between data-taking periods, \KS
reconstruction categories and the different final states are assumed to
factorise (in an analogous way to the widths of the signal distributions),
leaving six free parameters.
This assumption, as well as the choice of the linear model, are considered
as sources of systematic uncertainties. 

The fit results for each BDT optimisation, combining all data-taking
periods, are displayed in \figstwo{fitLoose2011}{fitTight2011}.
The separate plots for the individual data-taking periods are shown in
Figs.~\ref{fig:FitResult:Loose:DD:2011}--\ref{fig:FitResult:Tight:LL:2012b}
in \refapp{fit-results-individual-categories}.
Table~\ref{tab : FitResultsFused} shows the signal yields for each mode
summed over all data-taking periods and \KS reconstruction categories, along with a
weighted sum of efficiencies.
The fitted yields of each decay mode for each of the three data-taking
periods and two \KS reconstruction categories are given in
Appendix~\ref{sec : fit-results-individual-categories}.
Statistical correlations between the signal yields are below $10\%$ in all cases
and are neglected.
For the suppressed modes, the combinatorial background is negligible
in the high invariant-mass region for the \KsPiPi and \KsKK final states,
leading to a small systematic uncertainty related to the assumptions used
to fit this component.
In order to determine the significance of the \BstoKsKK signal, likelihood
profiles are constructed for the \BstoKsKK yield in each fit category,
taking into account systematic uncertainties.
The profiles are constructed from fits where the shape parameters of the
\BstoKsKK signal are fixed to the values obtained from the nominal fit,
which allows the change in the fit likelihood to be interpreted using
Wilks' theorem~\cite{Wilks:1938dza}.
Combining these profiles yields a significance of $2.5\,\sigma$.  

\begin{table}[t]
  \caption{Signal yields obtained from the simultaneous fit to the data.
    The yields are the sum of those obtained in the three data-taking periods when
    fitting the data sample selected using the BDT optimisation chosen for
    the given decay mode.
    The uncertainties are statistical only.
    The average selection efficiencies, described in
    Sec.~\ref{sec:Efficiencies}, are also shown for each decay mode
    together with the corresponding total uncertainty due to the limited
    simulation sample size and systematic effects in their determination.
  }
  \label{tab : FitResultsFused}
  \begin{center}
    \begin{tabular}{l | >{\hfill} p{0.9cm}@{$\,\pm\,$}p{0.5cm}  >{\hfill} p{1.25cm}@{$\,\pm\,$}p{1.25cm} | >{\hfill} p{0.9cm}@{$\,\pm\,$}p{0.5cm}  >{\hfill} p{1.25cm}@{$\,\pm\,$}p{1.25cm} }
            & \multicolumn{4}{c|}{\DD}                                         & \multicolumn{4}{c}{\LL}                                         \\
      Decay       & \multicolumn{2}{c}{Yield} & \multicolumn{2}{c|}{Efficiency (\%)} & \multicolumn{2}{c}{Yield} & \multicolumn{2}{c}{Efficiency (\%)} \\
      \hline
      \BdtoKsPiPi & $2766$ & $66$              & $0.0447$ & $0.0039$                  & $1411$ & $45$              & $0.0168$ & $0.0015$                 \\
      \BdtoKsKPi  &  $261$ & $24$              & $0.0340$ & $0.0031$                  & $160$  & $17$               & $0.0120$ & $0.0012$                 \\
      \BdtoKsKK   & $1133$ & $39$              & $0.0300$ & $0.0035$                  & $685$ & $29$              & $0.0142$ & $0.0017$                 \\
      \BstoKsPiPi &  $146$ & $19$              & $0.0359$ & $0.0030$                  & $74$  & $11$               & $0.0127$ & $0.0011$                 \\
      \BstoKsKPi  & $1100$ & $41$              & $0.0387$ & $0.0035$                  & $568$ & $28$              & $0.0146$ & $0.0013$                 \\
      \BstoKsKK   &  $12$  & $6$               & $0.0282$ & $0.0023$                  & $7$   & $4$               & $0.0094$ & $0.0013$                 \\
    \end{tabular}
  \end{center}
\end{table}

\paperFitResults{Loose}{2011}
\paperFitResults{Tight}{2011}

%% file: efficiencies.tex
\section{Determination of the efficiencies} 
\label{sec:Efficiencies}

The measurements of the branching fractions of the \BdstoKshhp decays relative to the well established  \BdtoKsPiPi decay 
mode proceed according to
\begin{eqnarray}      
\frac{\BF(\BdstoKshhp)}{\BF(\BdtoKsPiPi)} & = &
\frac{\eps^{\rm sel}_{\BdtoKsPiPi}}{\eps^{\rm sel}_{\BdstoKshhp}}
\frac{N_{\BdstoKshhp} }{ N_{\BdtoKsPiPi}} \frac{f_d}{f_{d,s}}\,,
\end{eqnarray}  
where $\eps^{\rm sel}$ is the selection efficiency (which includes geometrical acceptance,
 reconstruction, selection, trigger and particle identification
components), $N$ is the fitted signal yield, and $f_d$ and $f_s$ are the 
hadronisation fractions of a \bquark quark into a \Bz and \Bs 
meson, respectively. The ratio \fsfdinline has been precisely determined by the
 \lhcb experiment from hadronic and semileptonic measurements to be
$\fsfdinline = 0.259 \pm 0.015$~\cite{fsfd}. 
Since the \CP content of the three \Bs decays is currently unknown, the calculation
 of the corresponding efficiencies assumes an effective lifetime of $1/\Gs$, where
 $\Gs$ is the average width of the two \CP-eigenstates of the \Bs meson.
The effect of varying the decay width by $\pm\DGs/2$, where $\DGs$ is the width
difference between the two \Bs\ \CP-eigenstates, results in relative changes to
the average efficiency of $\mp4\%$.

Three-body decays are, in general, composed of several quasi-two-body decays and nonresonant
contributions, all of them possibly interfering.
The signal reconstruction, selection and trigger efficiencies also vary over the phase space.
Hence, both the distribution of the signal events and the variation of the
efficiency over the Dalitz plot~\cite{Dalitz:1953cp} must be
determined in order to calculate the efficiency-corrected yield.
In this analysis, efficiencies are determined for each decay mode from
simulated signal samples in bins of the ``square Dalitz
plot''~\cite{Aubert:2005sk}, where the usual Dalitz-plot coordinates have
been transformed in order to map a rectangular space.
The edges of the phase space are spread out in the square Dalitz plot,
which permits a more precise modelling of the efficiency in the regions
where it varies the most and where most of the signal candidates are
expected.
The square Dalitz-plot distribution of each signal mode is determined from
the data using the \sPlot\ technique~\cite{Pivk:2004ty}, using the \Kshh
invariant mass as the discriminating variable.
The distributions of signal events on the Dalitz plot, as obtained using this technique,
are shown in Appendix~\ref{sec : Dalitz-plot-sWeights}.
The efficiency-corrected yields are calculated as:
\begin{equation}
N^\mathrm{corr}_{\BdstoKshhp} = \sum_i^N \frac{w_i}{\varepsilon_i} \,,
\end{equation}
where $w_i$ is the \sPlot\ weight and $\varepsilon_i$ is the efficiency for
event $i$, and the sum includes all events in the fitted data sample.
The average efficiency for each decay mode:
\begin{equation}
\overline{\varepsilon} = \frac{N_{\BdstoKshhp}}{N^\mathrm{corr}_{\BdstoKshhp}} \,,
\end{equation}
where $N_{\BdstoKshhp}$ is the fitted signal yield, is given for each
signal decay in Table~\ref{tab : FitResultsFused}.
They are presented for each data-taking period and \KS reconstruction
category in \tabs{FitResults2011}{FitResults2012b} in Appendix~\ref{sec : fit-results-individual-categories}.
Their relative uncertainties due to the finite size of the simulated event
samples vary from 1\% to 20\%.

The imperfections of the simulation are corrected for in several respects.
Inaccuracies of the tracking efficiency in the simulation are mitigated by
weighting the \hadp and \hadprimm tracks by a correction factor obtained
from a data calibration sample~\cite{LHCb-DP-2013-002}.
An analogous correction is applied for the \KS tracking and vertex reconstruction efficiency.
A control data sample of $\Dstarp \to \Dz (\to \Km\pip) \pi^+_s$ decays,
where $\pi^+_s$ indicates a slow pion,
is used to quantify the differences between the data and simulation hardware
trigger stage for pions and kaons, independently for positive and negative 
hadron charges, as a function of \pt~\cite{LHCb-DP-2012-004}.
Corrections to account for differences between data and simulation related
to tracking efficiency are $\order(1\%)$, while those related to
trigger efficiency can be $\order(10\%)$, depending on the position on the
Dalitz plot.
The uncertainties attached to these various corrections are propagated to
the branching fraction measurements as systematic uncertainties and are
further discussed in Sec.~\ref{sec : systematics}.

The particle identification efficiencies and misidentification rates are
determined from simulated signal decays on an event-by-event basis using a
data-driven calibration to match the kinematic properties of the tracks in
the decay of interest.
A weighting procedure is performed in bins of \ptot, $\eta$ and event
multiplicity, accounting for kinematic correlations between the tracks.
Calibration tracks are taken from $\Dstarp \to \Dz \pi^+_s$ decays where
the \Dz~decays to the Cabibbo-favoured $\Km\pip$ final state.
In this case, the charge of the soft pion $\pi^+_s$ provides the
kaon or pion identity of the tracks. 
The dependence of the PID efficiency over the Dalitz plot induced by the
variations of PID performance with the track kinematics is included in the
procedure described above.
This calibration is performed using samples from the same data-taking
period, accounting for the variation in the performance of the Cherenkov
detectors over time.

%% file: systematics.tex
\section{Systematic uncertainties}
\label{sec : systematics}

Most of the systematic uncertainties are eliminated or greatly reduced by
normalising the branching fraction measurements to the \BdtoKsPiPi mode.
A summary of the contributions, expressed as relative uncertainties, is
given in \tab{syst_BR_COMPACT}, including the uncertainty in the
measurement of \fsfdinline~\cite{fsfd}.
A detailed breakdown of systematic uncertainties per data-taking period and
\KS reconstruction category is provided in Tables~\ref{tab : syst_BR_DD}
and~\ref{tab : syst_BR_LL} in Appendix~\ref{sec : systematics-breakdown}.
The dominant contributions arise from the modelling of the combinatorial
background shape in the invariant mass fit and from the determination of
the efficiency of the hardware trigger.

\renewcommand{\arraystretch}{1.3}

\begin{table}[t]
\caption{
Summary of the systematic uncertainties on the ratio of branching fractions.
A weighted average of the two \KS reconstruction categories and three data-taking periods is performed.
The values quoted for the individual contributions, which are illustrative
of the hierarchy between sources of systematic uncertainty, each result
from a weighted average in which the other systematic uncertainty
contributions are disregarded.
The total uncertainty is the weighted average including all contributions.
All uncertainties are relative and are quoted as percentages.
}
\label{tab : syst_BR_COMPACT}
\begin{center}
\resizebox{1.0\textwidth}{!}{
\begin{tabular}{ l c | c c c c c }
Relative \BF & & $\frac{\BR({\BdtoKsKPi})}{\BR({\BdtoKsPiPi})}$ & $\frac{\BR({\BdtoKsKK})}{\BR({\BdtoKsPiPi})}$ & $\frac{\BR({\BstoKsPiPi})}{\BR({\BdtoKsPiPi})}$ & $\frac{\BR({\BstoKsKPi})}{\BR({\BdtoKsPiPi})}$ & $\frac{\BR({\BstoKsKK})}{\BR({\BdtoKsPiPi})}$ \\
\hline
Fit model & [\%]        & $\phz9.7$ & $2.1$ & $13.5$     & $4.7$ & $18.5$ \\
Selection & [\%]        & $\phz3.8$ & $1.9$ & $\phz3.3$  & $2.4$ & $\phz6.7$ \\
Tracking & [\%]         & $\phz0.2$ & $0.1$ & $\phz0.2$  & $0.1$ & $\phz0.3$ \\
Trigger & [\%]          & $\phz3.2$ & $5.0$ & $\phz6.8$  & $3.5$ & $12.6$ \\
PID & [\%]              & $\phz1.1$ & $1.1$ & $\phz1.1$  & $1.1$ & $\phz1.1$ \\
\hline
Total & [\%]            & $12.2$    & $6.0$ & $16.2$     & $6.5$ & $26.9$ \\
\fsfdinline\ & [\%]     & $\cdots$  & $\cdots$      & $\phz5.8$          & $5.8$         & $\phz5.8$ \\
\end{tabular}
}
\end{center}
\end{table}

\renewcommand{\arraystretch}{1.0}

\subsection{Fit model}
\label{sec : fitmodel syst}

The fit model relies on a number of assumptions, both in terms of the values of
parameters being taken from simulation and in terms of the choice of the functional
forms describing the various contributions.
In both cases, the uncertainties are evaluated using pseudoexperiments that
are generated from the alternative parameterisation and are fitted using both the
nominal and the alternative fit models.
The distribution of the difference in the value of a given parameter
determined in the two fit model is subsequently fitted with a Gaussian
function and the corresponding systematic uncertainty is assigned as the
sum in quadrature of the mean and the resolution of the Gaussian.

This procedure is employed for the fixed parameters of the signal, partially-reconstructed
and cross-feed backgrounds and for the functional forms used for the signal
and combinatorial background.
Due to the limited sizes of the simulated event samples used to parameterise
both the partially-reconstructed and cross-feed background shapes, the
uncertainty related to the fixed parameters also covers any reasonable
variation of the shape.
For the combinatorial background, the ratios of the slopes in different \KS
reconstruction categories and in different data-taking periods are
constrained to be the same for all final states.
Two alternative models are considered:
allowing independent ratios for each of the final states (testing the
assumption of the factorisation of the slope ratios) and using an
exponential model instead of the nominal linear one (testing the functional
form of the combinatorial shape).

Finally, in order to evaluate the impact of residual contributions from \Lb
decays that survive the proton PID veto described in Sec.~\ref{sec:Selection},
fits to data are performed including a model for this contribution.
As these fits show negligible difference to the nominal model,
no systematic uncertainty is assigned.

The total fit model systematic uncertainty is given by the sum in
quadrature of all the contributions and is mostly dominated by the
combinatorial background model uncertainty.
Some uncertainties are fully correlated among the different data samples that
are averaged and are treated as such in the uncertainty propagation. 
The correlated fit model systematics include uncertainties due to the fit biases and combinatorial and signal shapes. 
The combination of these contributions is shown in Table~\ref{tab : syst_BR_COMPACT} as ``Fit model'', while they are referred to in Tables~\ref{tab : syst_BR_DD} and~\ref{tab : syst_BR_LL} as ``Fit model (corr.)'' and ``Fit model (uncorr.)'', respectively.
 
\subsection{Selection and trigger efficiencies}
\label{sec : eff syst}

The accuracy of the efficiency determination is limited in most cases by the finite size of the samples of simulated signal events, duly propagated 
as a systematic uncertainty.  In addition, the effect related to the choice of binning for the square Dalitz plot is estimated from the  spread of the average 
efficiencies determined from several alternative binning schemes.
These two sources of uncertainties are detailed in Tables~\ref{tab :
syst_BR_DD} and~\ref{tab : syst_BR_LL}, and are labelled ``Selection
(statistics)'' and ``Selection (binning)'', respectively. 

As introduced in Sec.~\ref{sec:Efficiencies}, the sources of
uncertainties related to the imperfections of the tracking simulation are
two-fold:
the reconstruction of both long and downstream tracks and
the reconstruction of the \KS decay vertex (in particular for the
downstream category).
In both cases, the reconstruction efficiencies are determined from data
calibration samples and the simulated events are weighted to match the
performance measured in the data.
The uncertainties arising from the finite size of the calibration samples
are propagated to assign a systematic uncertainty.
  
Possible sources of systematic uncertainty related to the efficiency
estimation of the hardware trigger have been studied.
Two additional systematic uncertainties are assigned:
one related to the imperfect simulation of the rate of overlapping tracks
in the hadron calorimeter forming a single hadron trigger candidate and 
one related to the choice of the data calibration sample itself.
These two sources of uncertainties are detailed in
Tables~\ref{tab : syst_BR_DD} and~\ref{tab : syst_BR_LL} in
Appendix~\ref{sec : systematics-breakdown}, labelled ``Trigger (overlap)''
and ``Trigger (calib.~sample)'', respectively.
For the first source, the systematic uncertainty is estimated as the difference
between the trigger efficiency correction with and without the overlapping cluster corrections.
For the latter source of uncertainty, the correction factors have been
determined from a sample of reconstructed $\Bz\to\jpsi\Kp\pim$ events.
Twice the difference between the correction factors determined from the two
calibration samples is taken as the estimate of the associated systematic
uncertainty.

The uncertainties due to the choice of the binning of the square Dalitz
plot used to produce the efficiency maps and those related to the hardware
trigger calibration samples are treated as correlated among the different
data samples split by year of data taking.

\subsection{Particle identification efficiencies}
\label{sec : pid syst}

The procedure to evaluate the efficiencies of the PID selections uses calibration tracks that differ from the signal tracks in terms of their kinematic distributions.
While the binning procedure attempts to mitigate these differences, there could be some remaining systematic effect. This is addressed by considering different 
ensembles of kinematical binnings to determine the efficiency. An overall 1\%  systematic uncertainty is assigned to quantify any bias due to the procedure. 
The statistical uncertainties originating from the finite sample sizes are added in quadrature.

%% file: results.tex
\section{Results and conclusion} 
\label{sec:Results}

The decay modes \BdstoKshhp have been analysed using a dataset,
corresponding to an integrated luminosity of 3.0\invfb recorded by the LHCb
detector at a centre-of-mass energy of 7\tev and 8\tev.
The branching fraction of each decay is measured relative to that of \BdtoKsPiPi.
The ratios of branching fractions are determined independently for the two
\KS reconstruction categories and three data-taking periods and then
combined by performing a $\chi^2$ fit.
The corresponding covariance matrix includes the statistical and systematic
uncertainties.
A $100\%$ linear correlation factor is assumed for the correlated
systematic uncertainties.
Good agreement is obtained among all determinations from each data-taking period and \KS reconstruction category. 
The results obtained from the combination are
\begin{eqnarray*}
\nonumber
\frac{\Br{\BdtoKsKPi}} {\Br{\BdtoKsPiPi}}  & = & 0.123 \pm 0.009 \; \stat \; \pm 0.015 \; \syst \,,\\ 
\nonumber                                              
\frac{\Br{\BdtoKsKK}}  {\Br{\BdtoKsPiPi}}  & = & 0.549 \pm 0.018 \; \stat \; \pm 0.033 \; \syst \,,\\
\nonumber                                              
\frac{\Br{\BstoKsPiPi}}{\Br{\BdtoKsPiPi}}  & = & 0.191 \pm 0.027 \; \stat \; \pm  0.031 \; \syst \; \pm 0.011 \; (f_s/f_d) \,, \\
\nonumber  
\frac{\Br{\BstoKsKPi}} {\Br{\BdtoKsPiPi}}  & = & 1.70\phantom{0} \pm 0.07\phantom{0} \; \stat \; \pm  0.11\phantom{0} \; \syst \; \pm 0.10\phantom{0} \; (f_s/f_d) \,,\\
\nonumber    
\frac{\Br{\BstoKsKK}}{\Br{\BdtoKsPiPi}}    & = & 0.026 \pm 0.011 \; \stat \; \pm  0.007 \; \syst \; \pm 0.002 \; (f_s/f_d) \,.
\end{eqnarray*}  
All measurements of branching fractions are in good agreement with the earlier \lhcb
determinations~\cite{LHCb-PAPER-2013-042}, which they supersede.
The measurement of the relative branching fractions of \BdtoKsKPi and
\BdtoKsKK are consistent with the world average results~\cite{HFAG,PDG2016}. 

The significance of the measured signal yield for the
decay \BstoKsKK, including systematic uncertainties, is $2.5$ standard
deviations.
A 90\% confidence level (C.L.) interval for the corresponding
branching fraction relative to that of \BdtoKsPiPi is derived, following
the approach of Feldman--Cousins~\cite{Feldman:1997qc}
\begin{eqnarray*}
\nonumber
\frac{\Br{\BstoKsKK}}{\Br{\BdtoKsPiPi}}    \in  [0.008 - 0.051] \mathrm{~at~90\%~C.L.}
\end{eqnarray*}  
Using the world average value omitting the previous LHCb result, ${\cal B}(\BdtoKzPiPi) = (4.96 \pm 0.20) \times 10^{-5}$~\cite{HFAG,PDG2016}, the measured time-integrated branching fractions are 
\begin{eqnarray}
\nonumber
\Brr{\BdtoKzKPi}  &=& \phantom{0}(6.1 \pm 0.5  \pm  0.7 \pm 0.3)\times10^{-6} \,, \\
\nonumber
\Brr{\BdtoKzKK}   &=& (27.2 \pm 0.9 \pm 1.6 \pm 1.1)\times10^{-6} \,,\\
\nonumber
\Brr{\BstoKzPiPi} &=& \phantom{0}(9.5 \pm 1.3 \pm 1.5 \pm 0.4)\times10^{-6} \,,\\
\nonumber
\Brr{\BstoKzKPi}  &=& (84.3 \pm 3.5 \pm 7.4 \pm 3.4)\times10^{-6} \,,\\
\nonumber
\Brr{\BstoKzKK}   &\in& [0.4 - 2.5] \times10^{-6} \; \mathrm{~at~90\%~C.L.} \,,
\end{eqnarray}  
where the first uncertainty is statistical, the second systematic and the last due to the uncertainty on ${\cal B}(\BdtoKzPiPi)$.
These results are in agreement with the available predictions for these channels~\cite{Cheng:2013dua,Cheng:2014uga,Li:2014fla,Li:2014oca}.

The first Dalitz-plot analyses by the \lhcb experiment of the dominant
decays (\BdtoKsPiPi, \BstoKsKPi, and \BdtoKsKK) are the next
step of the physics programme introduced in this work.
These studies will follow and benefit from the selection methods developed
for this analysis.

%% file: acknowledgements.tex
\section*{Acknowledgements}

\noindent We express our gratitude to our colleagues in the CERN
accelerator departments for the excellent performance of the LHC. We
thank the technical and administrative staff at the LHCb
institutes. We acknowledge support from CERN and from the national
agencies: CAPES, CNPq, FAPERJ and FINEP (Brazil); MOST and NSFC (China);
CNRS/IN2P3 (France); BMBF, DFG and MPG (Germany); INFN (Italy); 
NWO (The Netherlands); MNiSW and NCN (Poland); MEN/IFA (Romania); 
MinES and FASO (Russia); MinECo (Spain); SNSF and SER (Switzerland); 
NASU (Ukraine); STFC (United Kingdom); NSF (USA).
We acknowledge the computing resources that are provided by CERN, IN2P3 (France), KIT and DESY (Germany), INFN (Italy), SURF (The Netherlands), PIC (Spain), GridPP (United Kingdom), RRCKI and Yandex LLC (Russia), CSCS (Switzerland), IFIN-HH (Romania), CBPF (Brazil), PL-GRID (Poland) and OSC (USA). We are indebted to the communities behind the multiple open 
source software packages on which we depend.
Individual groups or members have received support from AvH Foundation (Germany),
EPLANET, Marie Sk\l{}odowska-Curie Actions and ERC (European Union), 
Conseil G\'{e}n\'{e}ral de Haute-Savoie, Labex ENIGMASS and OCEVU, 
R\'{e}gion Auvergne (France), RFBR and Yandex LLC (Russia), GVA, XuntaGal and GENCAT (Spain), Herchel Smith Fund, The Royal Society, Royal Commission for the Exhibition of 1851 and the Leverhulme Trust (United Kingdom).

%% file: appendix.tex
\clearpage

{\noindent\normalfont\bfseries\Large Appendices}

\appendix
\FloatBarrier
\section{Fit results by category}
\label{sec : fit-results-individual-categories}

Signal yields and efficiencies for the different decays, data-taking periods and \KS reconstruction
categories are shown for each of the two BDT optimisation points in
\tabs{FitResults2011}{FitResults2012b}.
Fit results for the different data-taking periods and \KS reconstruction
categories are shown for each of the two BDT optimisation points in
Figs.~\ref{fig:FitResult:Loose:DD:2011}--\ref{fig:FitResult:Tight:LL:2012b}.

\begin{table}[htb]
  \paperCaption{2011}
  \begin{center}
    \begin{tabular}{l | >{\hfill} p{0.9cm}@{$\,\pm\,$}p{0.5cm}  >{\hfill} p{1.25cm}@{$\,\pm\,$}p{1.25cm} | >{\hfill} p{0.9cm}@{$\,\pm\,$}p{0.5cm}  >{\hfill} p{1.25cm}@{$\,\pm\,$}p{1.25cm} }
            & \multicolumn{4}{c|}{\DD}                                         & \multicolumn{4}{c}{\LL}                                         \\
      Decay       & \multicolumn{2}{c}{Yield} & \multicolumn{2}{c|}{Efficiency (\%)} & \multicolumn{2}{c}{Yield} & \multicolumn{2}{c}{Efficiency (\%)} \\
      \hline
      \BdtoKsPiPi & $803$ & $36$              & $0.0488$ & $0.0093$                  & $471$ & $27$              & $0.0188$ & $0.0036$                 \\
      \BdtoKsKK   & $281$ & $19$              & $0.0292$ & $0.0063$                  & $222$ & $17$              & $0.0157$ & $0.0034$                 \\
      \BstoKsKPi  & $333$ & $23$              & $0.0361$ & $0.0064$                  & $207$ & $16$              & $0.0148$ & $0.0025$                 \\
      \BdtoKsKPi  &  $76$  & $13$              & $0.0322$ & $0.0063$                  & $50$  & $9$                 & $0.0174$ & $0.0034$                 \\
      \BstoKsPiPi &  $43$  & $10$              & $0.0316$ & $0.0051$                  & $21$  & $8$                 & $0.0160$ & $0.0025$                 \\
      \BstoKsKK   &  $5$   & $3$                 & $0.0244$ & $0.0052$                  & $4$   & $3$                   & $0.0129$ & $0.0029$                 \\
    \end{tabular}
  \end{center}
\end{table} 

\begin{table}[htb]
  \paperCaption{2012a}
  \begin{center}
    \begin{tabular}{l | >{\hfill} p{0.9cm}@{$\,\pm\,$}p{0.5cm}  >{\hfill} p{1.25cm}@{$\,\pm\,$}p{1.25cm} | >{\hfill} p{0.9cm}@{$\,\pm\,$}p{0.5cm}  >{\hfill} p{1.25cm}@{$\,\pm\,$}p{1.25cm} }
      & \multicolumn{4}{c|}{\DD}                                         & \multicolumn{4}{c}{\LL}                                         \\
      Decay       & \multicolumn{2}{c}{Yield} & \multicolumn{2}{c|}{Efficiency (\%)} & \multicolumn{2}{c}{Yield} & \multicolumn{2}{c}{Efficiency (\%)} \\
      \hline
      \BdtoKsPiPi & $553$ & $30$              & $0.0423$ & $0.0059$                  & $286$ & $20$              & $0.0166$ & $0.0023$                 \\
      \BdtoKsKK   & $181$ & $15$              & $0.0263$ & $0.0052$                  & $119$ & $12$              & $0.0149$ & $0.0029$                 \\
      \BstoKsKPi  & $205$ & $18$              & $0.0395$ & $0.0060$                  & $99$  & $13$              & $0.0155$ & $0.0023$                 \\
      \BdtoKsKPi  & $63$  & $11$              & $0.0306$ & $0.0047$                  & $45$  & $10$              & $0.0143$ & $0.0022$                 \\
      \BstoKsPiPi & $17$  & $8$               & $0.0493$ & $0.0068$                  & $15$  & $6$               & $0.0145$ & $0.0021$                 \\
      \BstoKsKK   & $2$   & $3$               & $0.0290$ & $0.0039$                  & $1$   & $2$               & $0.0092$ & $0.0030$                 \\
    \end{tabular}
  \end{center}
\end{table} 

\begin{table}[htb]
  \paperCaption{2012b}
  \begin{center}
    \begin{tabular}{l  | >{\hfill} p{0.9cm}@{$\,\pm\,$}p{0.5cm}  >{\hfill} p{1.25cm}@{$\,\pm\,$}p{1.25cm} | >{\hfill} p{0.9cm}@{$\,\pm\,$}p{0.5cm}  >{\hfill} p{1.25cm}@{$\,\pm\,$}p{1.25cm} }
      & \multicolumn{4}{c|}{\DD}                                         & \multicolumn{4}{c}{\LL}                                         \\
      Decay       & \multicolumn{2}{c}{Yield} & \multicolumn{2}{c|}{Efficiency (\%)} & \multicolumn{2}{c}{Yield} & \multicolumn{2}{c}{Efficiency (\%)} \\
      \hline
      \BdtoKsPiPi & $1410$ & $46$              & $0.0455$ & $0.0063$                  & $654$ & $30$              & $0.0161$ & $0.0022$                 \\
      \BdtoKsKK   & $671$  & $30$              & $0.0395$ & $0.0076$                  & $344$ & $20$              & $0.0128$ & $0.0025$                 \\
      \BstoKsKPi  & $562$  & $29$              & $0.0401$ & $0.0059$                  & $262$ & $19$              & $0.0138$ & $0.0020$                 \\
      \BdtoKsKPi & $122$  & $17$              & $0.0402$ & $0.0056$                  & $65$  & $10$              & $0.0100$ & $0.0015$                 \\
      \BstoKsPiPi & $86$   & $14$              & $0.0335$ & $0.0045$                  & $38$  & $5$               & $0.0108$ & $0.0015$                 \\
      \BstoKsKK   & $5$    & $4$               & $0.0291$ & $0.0034$                  & $2$   & $2$               & $0.0083$ & $0.0017$                 \\
    \end{tabular}
  \end{center}
\end{table} 

\plotDataFitResults{Loose}{DD}{2011}
\plotDataFitResults{Loose}{DD}{2012a}
\plotDataFitResults{Loose}{DD}{2012b}

\plotDataFitResults{Loose}{LL}{2011}
\plotDataFitResults{Loose}{LL}{2012a}
\plotDataFitResults{Loose}{LL}{2012b}

\plotDataFitResults{Tight}{DD}{2011}
\plotDataFitResults{Tight}{DD}{2012a}
\plotDataFitResults{Tight}{DD}{2012b}

\plotDataFitResults{Tight}{LL}{2011}
\plotDataFitResults{Tight}{LL}{2012a}
\plotDataFitResults{Tight}{LL}{2012b}

\clearpage

\section{Breakdown of systematic uncertainties}
\label{sec : systematics-breakdown}

The full breakdown of the systematic uncertainties for each data-taking
period is given in \tabstwo{syst_BR_DD}{syst_BR_LL} for, respectively, the
\DD\ and \LL\ \KS reconstruction categories.

\begin{landscape}
\begin{table}[htb]
\caption{Systematic uncertainties on the ratios of branching fractions for \DD\ \KS reconstruction. All uncertainties are relative and are quoted as percentages.}
\label{tab : syst_BR_DD}
\begin{center}
\resizebox{1.1\textwidth}{!}{
\begin{tabular}{ l c | c c c c c}
Relative \BF (2011 sample) & & $\frac{\BR({\BdtoKsKPi})}{\BR({\BdtoKsPiPi})}$ & $\frac{\BR({\BdtoKsKK})}{\BR({\BdtoKsPiPi})}$ & $\frac{\BR({\BstoKsPiPi})}{\BR({\BdtoKsPiPi})}$ & $\frac{\BR({\BstoKsKPi})}{\BR({\BdtoKsPiPi})}$ & $\frac{\BR({\BstoKsKK})}{\BR({\BdtoKsPiPi})}$ \\
\hline
Fit model (uncorrelated) & [\%] &     $12.3$        & $ 3.2 $ & $ \phz7.4 $  & $ 4.4 $ & $    12.2 $ \\
Fit model (correlated) & [\%] &       $11.5$        & $ 2.4 $ & $ 11.6 $     & $ 5.8 $ & $ \phz4.8 $ \\
Selection (statistics) & [\%] &       $\phz4.1$     & $ 1.1 $ & $ \phz3.7 $  & $ 4.0 $ & $ \phz6.4 $ \\
Selection (binning) & [\%] &          $\phz2.5$     & $ 2.6 $ & $ \phz4.3 $  & $ 2.6 $ & $ \phz2.9 $ \\
Tracking & [\%] &                     $\phz0.1$     & $ 0.0 $ & $ \phz0.1 $  & $ 0.0 $ & $ \phz0.1 $ \\
Trigger (overlap) & [\%] &            $\phz0.1$     & $ 0.2 $ & $ \phz0.0 $  & $ 0.0 $ & $ \phz0.2 $ \\
Trigger (calibration sample) & [\%] & $\phz3.7$     & $ 3.7 $ & $ \phz6.1 $  & $ 4.2 $ & $ \phz7.5 $ \\
PID & [\%] &                          $\phz1.1$     & $ 1.1 $ & $ \phz1.1 $  & $ 1.1 $ & $ \phz1.1 $ \\
$f_s/f_d$ & [\%] &                     $\phz\cdots$  & $\cdots$ & $ \phz5.8 $ & $ 5.8 $ & $ \phz5.8 $ \\
\hline\hline
Relative \BF (2012a sample) & & $\frac{\BR({\BdtoKsKPi})}{\BR({\BdtoKsPiPi})}$ & $\frac{\BR({\BdtoKsKK})}{\BR({\BdtoKsPiPi})}$ & $\frac{\BR({\BstoKsPiPi})}{\BR({\BdtoKsPiPi})}$ & $\frac{\BR({\BstoKsKPi})}{\BR({\BdtoKsPiPi})}$ & $\frac{\BR({\BstoKsKK})}{\BR({\BdtoKsPiPi})}$ \\
\hline
Fit model (uncorrelated) & [\%] &     $ 4.2 $ & $ 0.8 $ & $ \phz4.5 $  & $ 4.6 $ & $ 58.1 $ \\
Fit model (correlated) & [\%] &       $ 7.8 $ & $ 1.6 $ & $ 15.8 $     & $ 4.8 $ & $ 16.7 $ \\  
Selection (statistics) & [\%] &       $ 4.1 $ & $ 1.8 $ & $ \phz2.3 $  & $ 1.3 $ & $ \phz6.1 $ \\
Selection (binning) & [\%] &          $ 4.1 $ & $ 3.7 $ & $ \phz2.7 $  & $ 3.4 $ & $ \phz2.0 $ \\
Tracking & [\%] &                     $ 0.1 $ & $ 0.1 $ & $ \phz0.1 $  & $ 0.1 $ & $ \phz0.2 $ \\
Trigger (overlap) & [\%] &            $ 0.0 $ & $ 0.2 $ & $ \phz0.3 $  & $ 0.0 $ & $ \phz0.0 $ \\
Trigger (calibration sample) & [\%] & $ 2.8 $ & $ 5.9 $ & $ \phz6.4 $  & $ 2.6 $ & $ 12.8 $ \\
PID & [\%] &                          $ 1.1 $ & $ 1.1 $ & $ \phz1.1 $  & $ 1.1 $ & $ \phz1.1 $ \\
$f_s/f_d$ & [\%] &                    $\cdots$ & $\cdots$ & $ \phz5.8 $ & $ 5.8 $ & $ \phz5.8 $ \\
\hline\hline
Relative \BF (2012b sample) & & $\frac{\BR({\BdtoKsKPi})}{\BR({\BdtoKsPiPi})}$ & $\frac{\BR({\BdtoKsKK})}{\BR({\BdtoKsPiPi})}$ & $\frac{\BR({\BstoKsPiPi})}{\BR({\BdtoKsPiPi})}$ & $\frac{\BR({\BstoKsKPi})}{\BR({\BdtoKsPiPi})}$ & $\frac{\BR({\BstoKsKK})}{\BR({\BdtoKsPiPi})}$ \\
\hline
Fit model (uncorrelated) & [\%] &     $ \phz9.8 $   & $ 1.8 $ & $ 10.1 $    & $ 2.1 $ & $ 26.4 $ \\
Fit model (correlated) & [\%] &       $ 10.9 $      & $ 1.9 $ & $ 11.2 $    & $ 4.6 $ & $ 28.0 $ \\
Selection (statistics) & [\%] &       $ \phz1.2 $   & $ 0.5 $ & $ \phz1.1 $ & $ 0.7 $ & $ \phz3.4 $ \\
Selection (binning) & [\%] &          $ \phz1.6 $   & $ 1.3 $ & $ \phz1.3 $ & $ 3.3 $ & $ \phz3.8 $ \\
Tracking & [\%] &                     $ \phz0.1 $   & $ 0.0 $ & $ \phz0.0 $ & $ 0.0 $ & $ \phz0.2 $ \\
Trigger (overlap) & [\%] &            $ \phz0.2 $   & $ 0.2 $ & $ \phz0.1 $ & $ 0.0 $ & $ \phz0.2 $ \\
Trigger (calibration sample) & [\%] & $ \phz2.8 $   & $ 5.9 $ & $ \phz6.4 $ & $ 2.6 $ & $ 12.8 $ \\
PID & [\%] &                          $ \phz1.1 $   & $ 1.1 $ & $ \phz1.1 $ & $ 1.1 $ & $ \phz1.1 $ \\
$f_s/f_d$ & [\%] &                     $\phz\cdots$ & $\cdots$ & $ \phz5.8 $ & $ 5.8 $ & $ \phz5.8 $ \\
\end{tabular}
}
\end{center}
\end{table}
\end{landscape}

\begin{landscape}
\begin{table}[htb]
\caption{Systematic uncertainties on the ratios of branching fractions for \LL\ \KS reconstruction. All uncertainties are relative and are quoted as percentages.}
\label{tab : syst_BR_LL}
\begin{center}
\resizebox{1.1\textwidth}{!}{
\begin{tabular}{ l c | c c c c c}
Relative \BF (2011 sample) & & $\frac{\BR({\BdtoKsKPi})}{\BR({\BdtoKsPiPi})}$ & $\frac{\BR({\BdtoKsKK})}{\BR({\BdtoKsPiPi})}$ & $\frac{\BR({\BstoKsPiPi})}{\BR({\BdtoKsPiPi})}$ & $\frac{\BR({\BstoKsKPi})}{\BR({\BdtoKsPiPi})}$ & $\frac{\BR({\BstoKsKK})}{\BR({\BdtoKsPiPi})}$ \\
\hline
Fit model (uncorrelated) & [\%] &     $ 9.7 $  & $ 0.9 $  & $ 6.7 $ & $ 1.6 $ & $ 32.7$ \\     
Fit model (correlated) & [\%] &       $ 8.9 $  & $ 1.9 $  & $ 9.8 $ & $ 5.9 $ & $ 20.2$ \\       
Selection (statistics) & [\%] &       $ 2.9 $  & $ 1.6 $  & $ 3.3 $ & $ 2.5 $ & $ \phz8.0 $ \\      
Selection (binning) & [\%] &          $ 3.2 $  & $ 1.0 $  & $ 2.6 $ & $ 1.4 $ & $ \phz2.9 $ \\         
Tracking & [\%] &                     $ 0.1 $  & $ 0.1 $  & $ 0.2 $ & $ 0.1 $ & $ \phz0.4 $ \\                    
Trigger (overlap) & [\%] &            $ 0.3 $  & $ 0.2 $  & $ 0.1 $ & $ 0.1 $ & $ \phz0.2 $ \\            
Trigger (calibration sample) & [\%] & $ 3.7 $  & $ 3.7 $  & $ 6.1 $ & $ 4.2 $ & $ \phz7.5 $ \\
PID & [\%] &                          $ 1.1 $  & $ 1.1 $  & $ 1.1 $ & $ 1.1 $ & $ \phz1.1 $ \\
$f_s/f_d$ & [\%] &                    $\cdots$ & $\cdots$ & $ 5.8 $ & $ 5.8 $ & $ \phz5.8 $ \\
\hline\hline
Relative \BF (2012a sample) & & $\frac{\BR({\BdtoKsKPi})}{\BR({\BdtoKsPiPi})}$ & $\frac{\BR({\BdtoKsKK})}{\BR({\BdtoKsPiPi})}$ & $\frac{\BR({\BstoKsPiPi})}{\BR({\BdtoKsPiPi})}$ & $\frac{\BR({\BstoKsKPi})}{\BR({\BdtoKsPiPi})}$ & $\frac{\BR({\BstoKsKK})}{\BR({\BdtoKsPiPi})}$ \\
\hline
Fit model (uncorrelated) & [\%] &     $ 5.2 $  & $ 1.8 $  & $ 5.4 $ & $ 4.3 $ & $ 42.9$ \\    
Fit model (correlated) & [\%] &       $ 8.9 $  & $ 1.9 $  & $ 9.8 $ & $ 5.9 $ & $ 20.2$ \\      
Selection (statistics) & [\%] &       $ 5.2 $  & $ 2.8 $  & $ 4.5 $ & $ 2.1 $ & $ 21.1$ \\      
Selection (binning) & [\%] &          $ 2.0 $  & $ 1.1 $  & $ 2.8 $ & $ 2.4 $ & $ \phz7.5 $ \\         
Tracking & [\%] &                     $ 0.3 $  & $ 0.1 $  & $ 0.3 $ & $ 0.2 $ & $ \phz0.9 $ \\                     
Trigger (overlap) & [\%] &            $ 0.2 $  & $ 0.0 $  & $ 0.2 $ & $ 0.1 $ & $ \phz0.5 $ \\            
Trigger (calibration sample) & [\%] & $ 2.8 $  & $ 5.9 $  & $ 6.4 $ & $ 2.6 $ & $ 12.8 $ \\ 
PID & [\%] &                          $ 1.1 $  & $ 1.1 $  & $ 1.1 $ & $ 1.1 $ & $ \phz1.1 $ \\
$f_s/f_d$ & [\%] &                    $\cdots$ & $\cdots$ & $ 5.8 $ & $ 5.8 $ & $ \phz5.8 $ \\
\hline\hline
Relative \BF (2012b sample) & & $\frac{\BR({\BdtoKsKPi})}{\BR({\BdtoKsPiPi})}$ & $\frac{\BR({\BdtoKsKK})}{\BR({\BdtoKsPiPi})}$ & $\frac{\BR({\BstoKsPiPi})}{\BR({\BdtoKsPiPi})}$ & $\frac{\BR({\BstoKsKPi})}{\BR({\BdtoKsPiPi})}$ & $\frac{\BR({\BstoKsKK})}{\BR({\BdtoKsPiPi})}$ \\
\hline
Fit model (uncorrelated) & [\%] &      $ 4.4 $  & $ 2.8 $  & $ 9.2 $ & $ 4.5 $ & $ 80.3$ \\    
Fit model (correlated) & [\%] &        $ 4.6 $  & $ 1.4 $  & $ 9.1 $ & $ 4.0 $ & $ 15.0$ \\      
Selection (statistics) & [\%] &        $ 4.0 $  & $ 1.0 $  & $ 2.8 $ & $ 1.6 $ & $ 11.0$ \\      
Selection (binning) & [\%] &           $ 1.6 $  & $ 1.5 $  & $ 1.2 $ & $ 1.7 $ & $ \phz8.8 $ \\         
Tracking & [\%] &                      $ 0.3 $  & $ 0.1 $  & $ 0.2 $ & $ 0.1 $ & $ \phz0.4 $ \\                     
Trigger (overlap) & [\%] &             $ 0.1 $  & $ 0.1 $  & $ 0.1 $ & $ 0.1 $ & $ \phz0.1 $ \\            
Trigger (calibration sample) & [\%] &  $ 2.8 $  & $ 5.9 $  & $ 6.4 $ & $ 2.6 $ & $ 12.8 $ \\
PID & [\%] &                           $ 1.1 $  & $ 1.1 $  & $ 1.1 $ & $ 1.1 $ & $ \phz1.1 $ \\
$f_s/f_d$ & [\%] &                     $\cdots$ & $\cdots$ & $ 5.8 $ & $ 5.8 $ & $ \phz5.8 $ \\
\end{tabular}
}
\end{center}
\end{table}
\end{landscape}

\section{Dalitz-plot distributions of signal events}
\label{sec : Dalitz-plot-sWeights}

\begin{figure}[!htbp]
  \begin{center}
    \plotOne{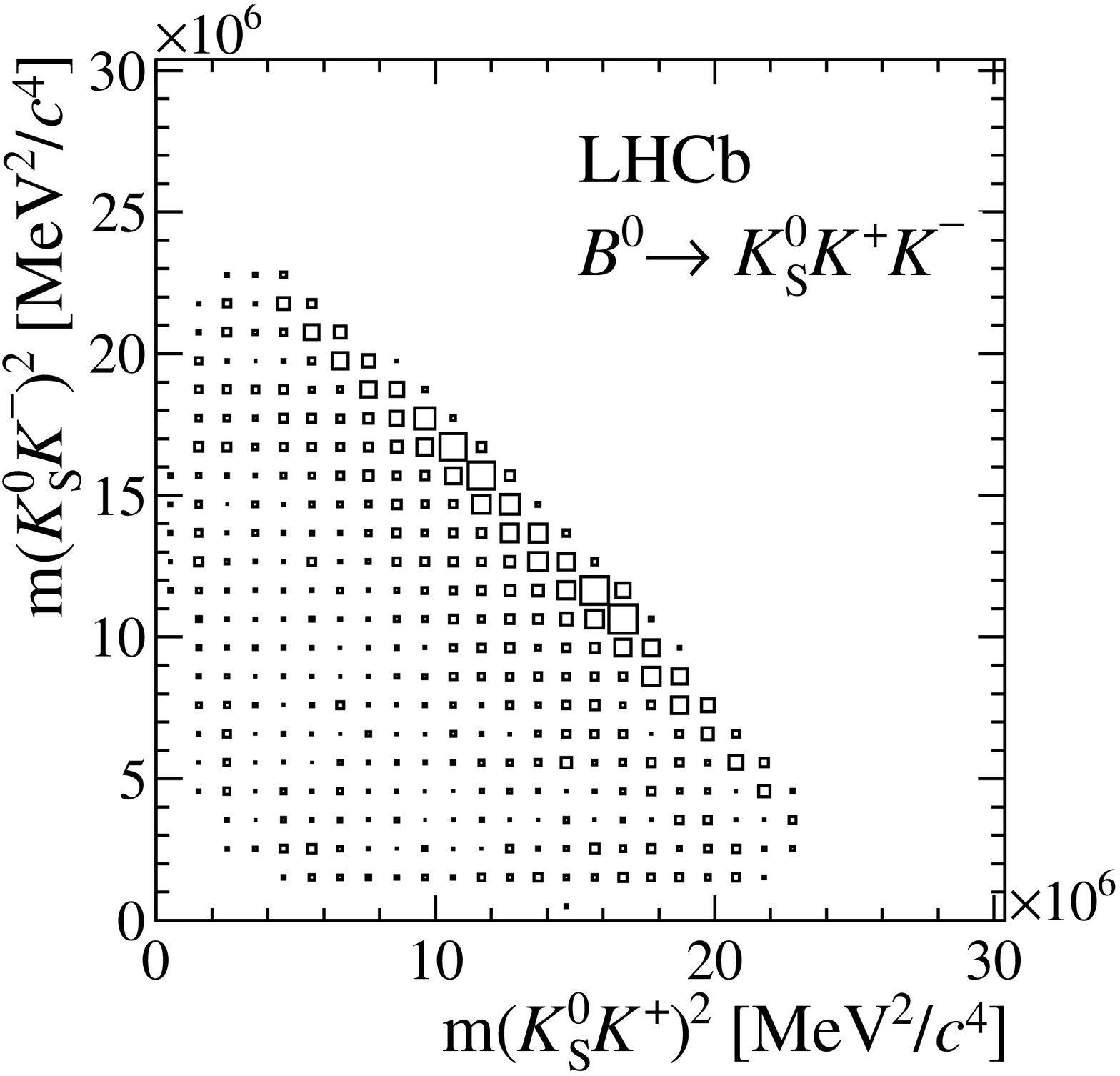}{0.35}
    \plotOne{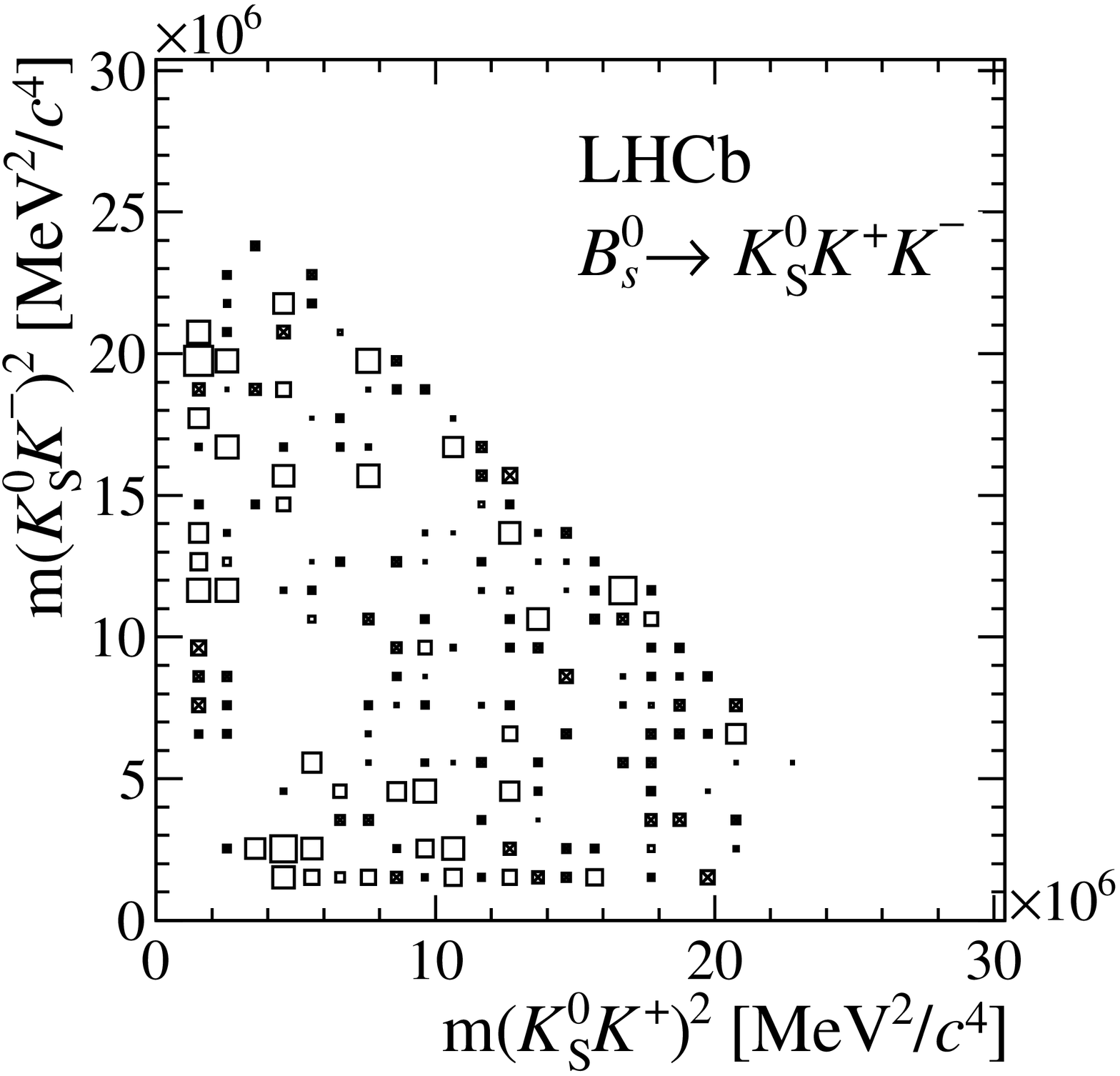}{0.35}\\
    \plotOne{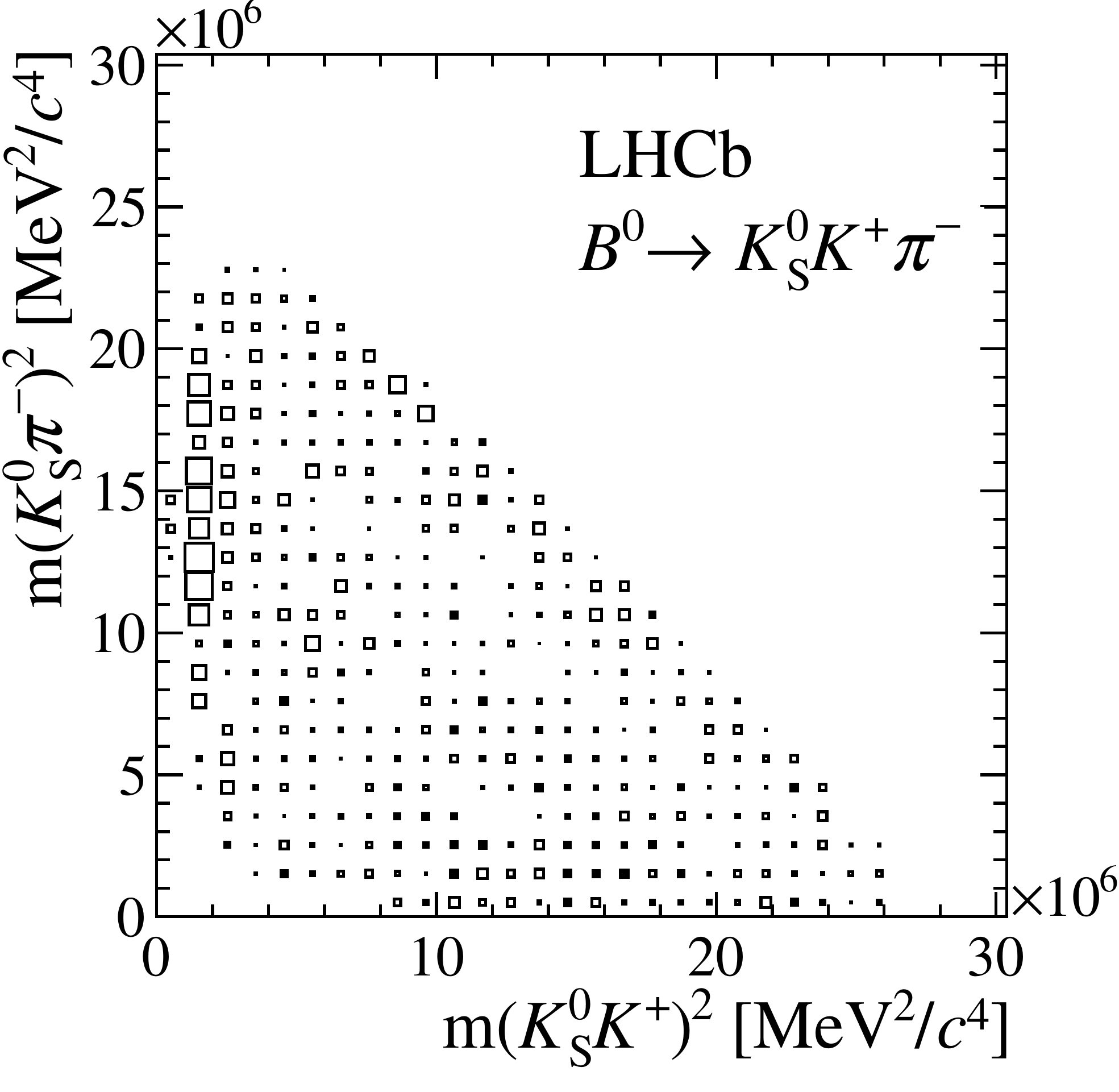}{0.35}
    \plotOne{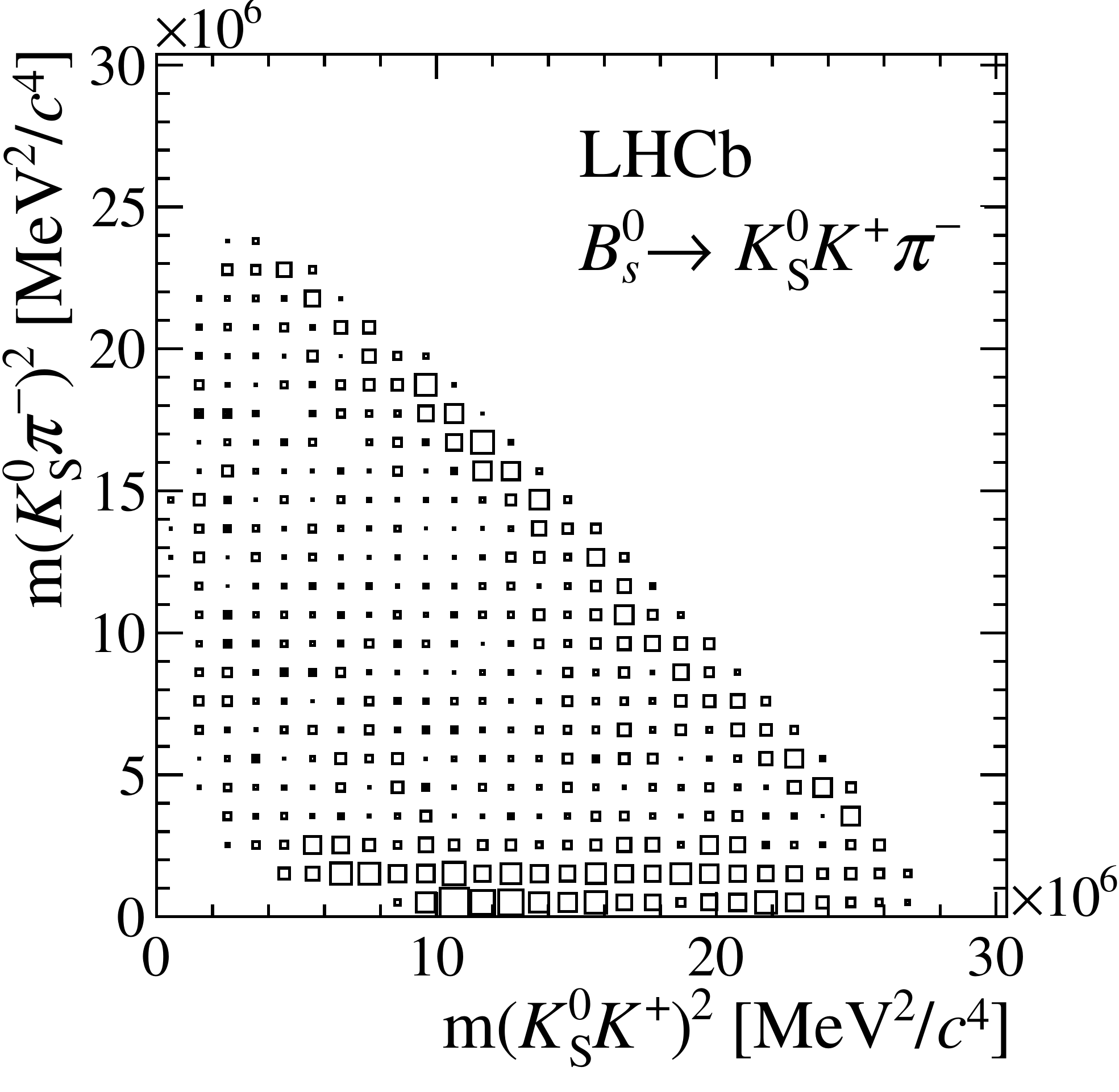}{0.35}\\
    \plotOne{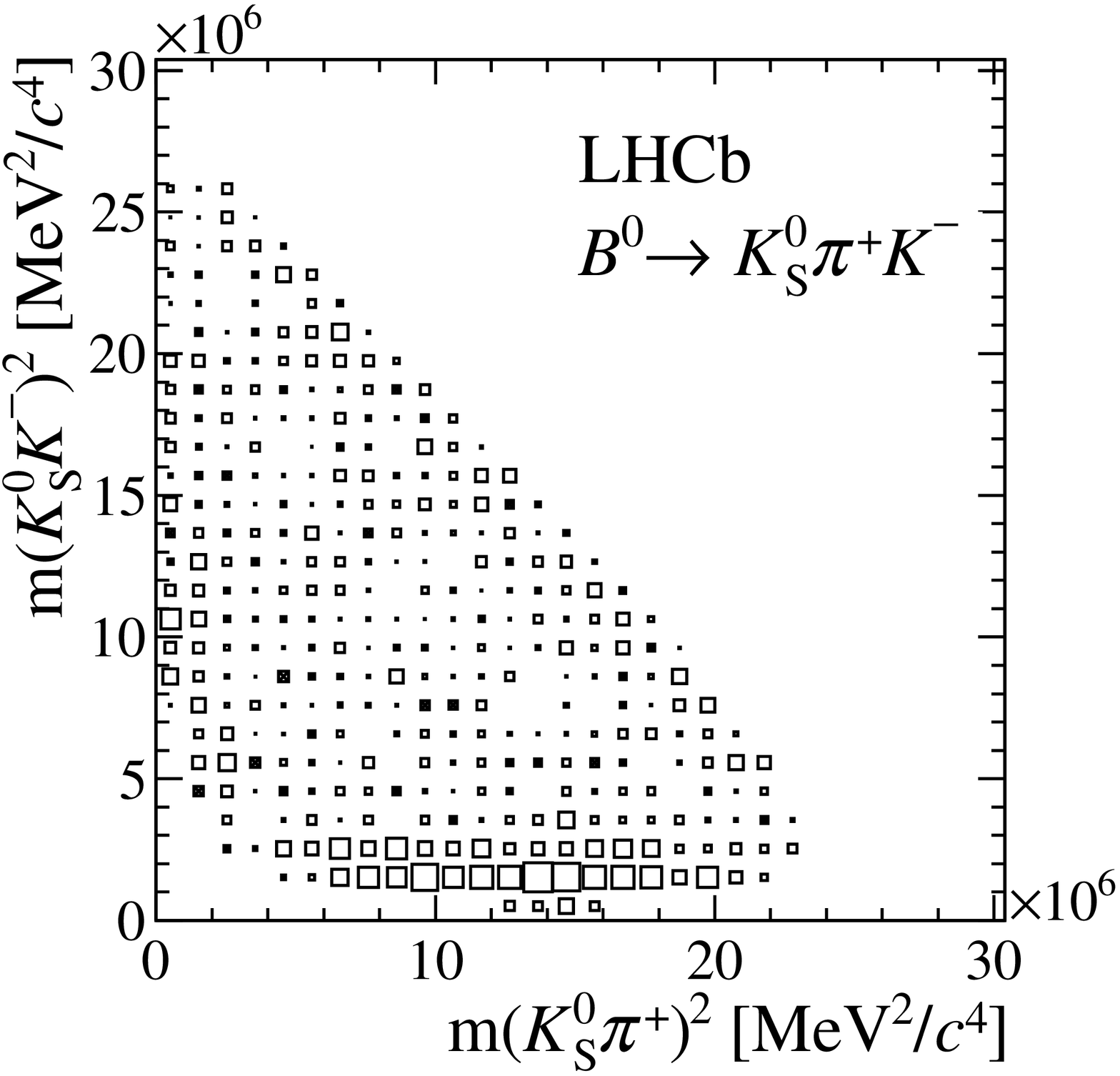}{0.35}
    \plotOne{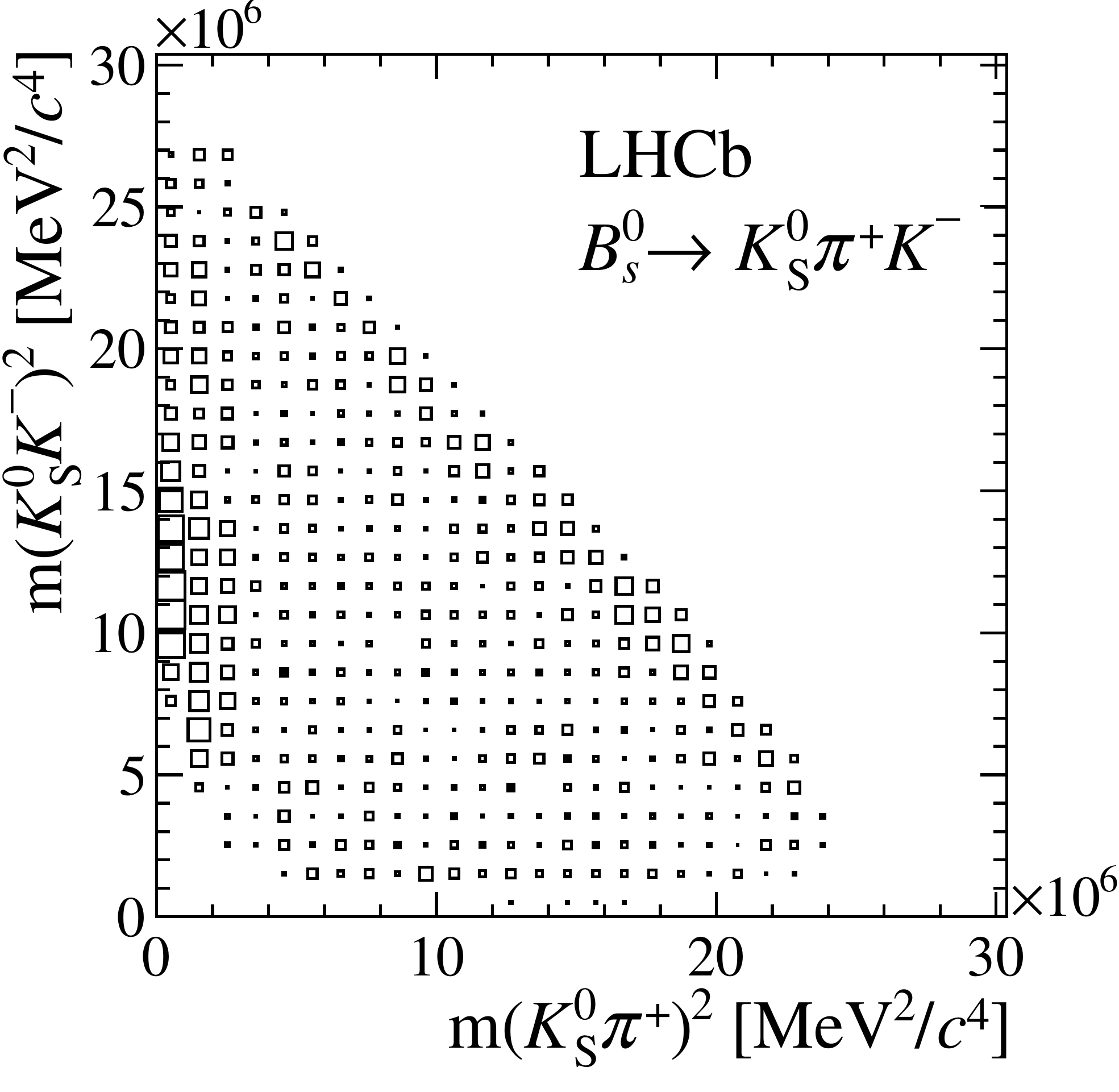}{0.35}\\
    \plotOne{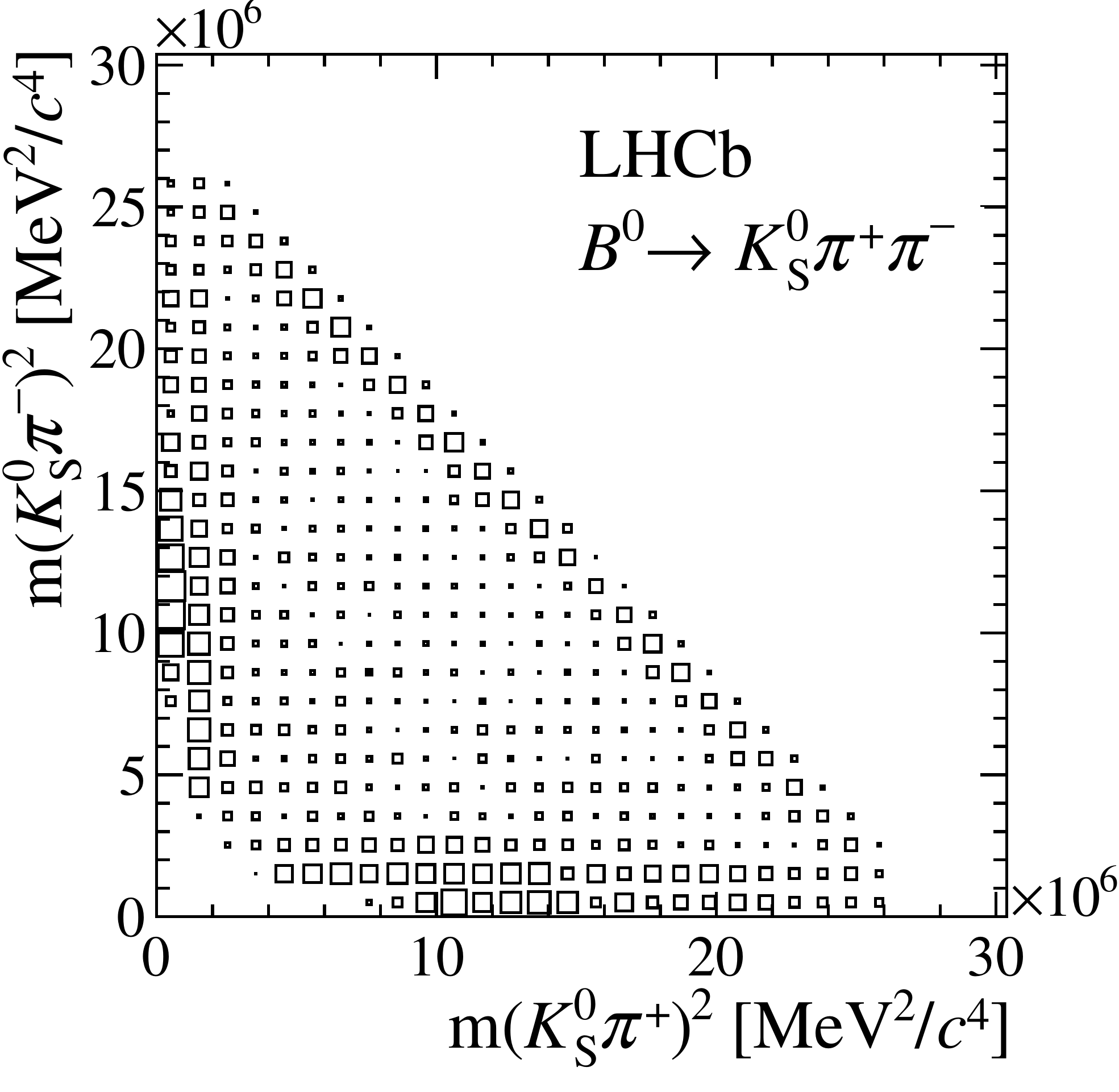}{0.35}
    \plotOne{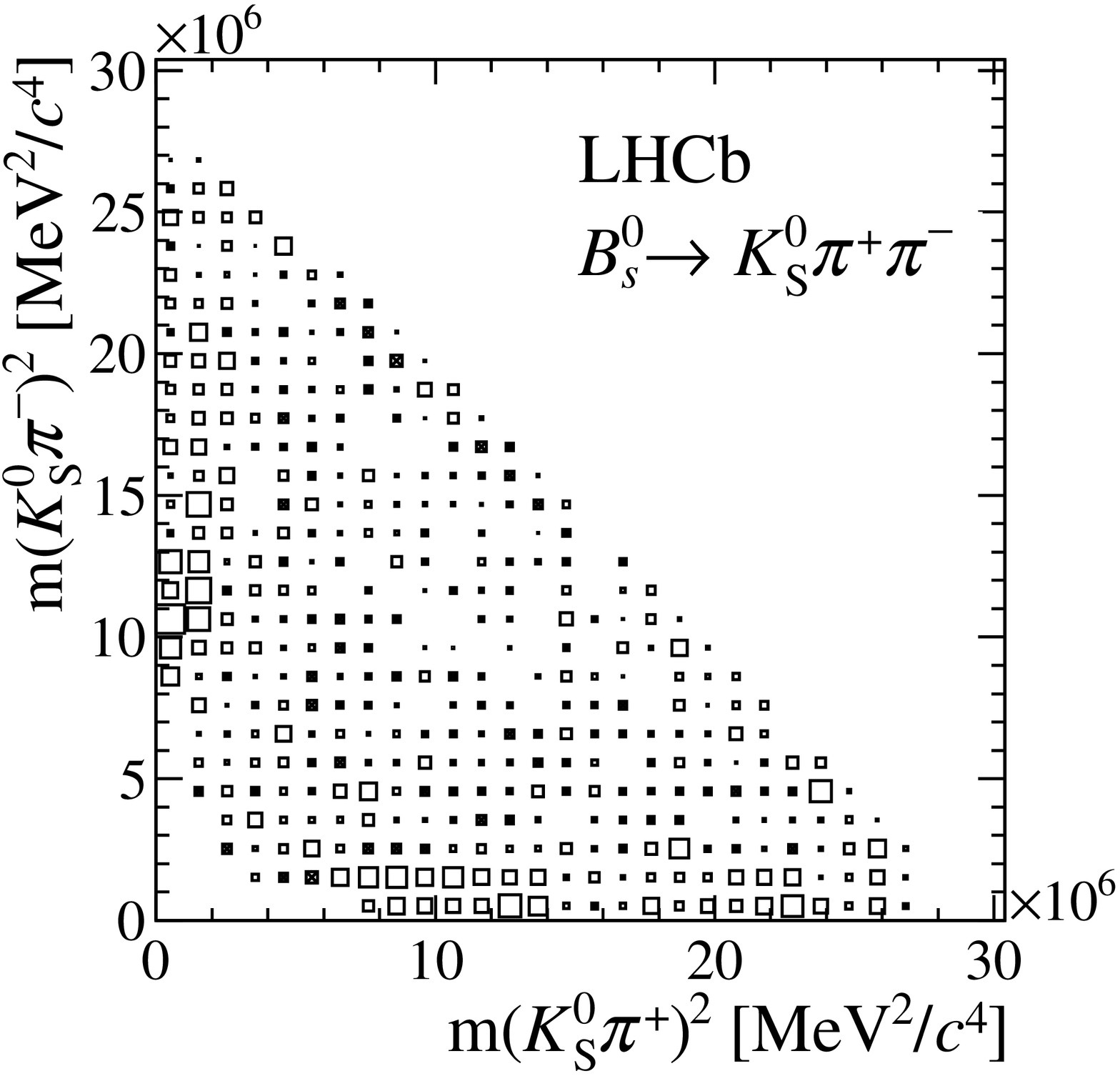}{0.35}\\
    \caption{Distribution of \sPlot\ weights in data.
      The \KsKK, \KsKpPim, \KsPipKm, and \KsPiPi final states are shown from top to bottom, with a \Bd parent on the left, and a \Bs on the right.
      All data-taking periods and \KS reconstruction categories are added.}
    \label{fig:FitResult:Splots:Dalitz}
  \end{center}
\end{figure}

Dalitz-plot distributions of signal events as extracted using the \sPlot\ technique
are shown for each signal mode are shown in Fig.~\ref{fig:FitResult:Splots:Dalitz}.

\clearpage

%% file: LHCb_Authorship_flat_07-Mar-2017.tex
\centerline{\large\bf LHCb collaboration}
\begin{flushleft}
\small
R.~Aaij$^{40}$,
B.~Adeva$^{39}$,
M.~Adinolfi$^{48}$,
Z.~Ajaltouni$^{5}$,
S.~Akar$^{59}$,
J.~Albrecht$^{10}$,
F.~Alessio$^{40}$,
M.~Alexander$^{53}$,
S.~Ali$^{43}$,
G.~Alkhazov$^{31}$
% ...

{\footnotesize \it
% ...
$ ^{5}$Clermont Universit{\'e}, Universit{\'e} Blaise Pascal, CNRS/IN2P3, LPC, Clermont-Ferrand, France\\
% ...
$ ^{10}$Fakult{\"a}t Physik, Technische Universit{\"a}t Dortmund, Dortmund, Germany\\
% ...
$ ^{31}$Petersburg Nuclear Physics Institute (PNPI), Gatchina, Russia\\
% ...
$ ^{39}$Universidad de Santiago de Compostela, Santiago de Compostela, Spain\\
$ ^{40}$European Organization for Nuclear Research (CERN), Geneva, Switzerland\\
% ...
$ ^{43}$Nikhef National Institute for Subatomic Physics, Amsterdam, The Netherlands\\
% ...
$ ^{48}$H.H. Wills Physics Laboratory, University of Bristol, Bristol, United Kingdom\\
% ...
$ ^{53}$School of Physics and Astronomy, University of Glasgow, Glasgow, United Kingdom\\
% ...
$ ^{59}$University of Cincinnati, Cincinnati, OH, United States\\
% ...
}
\end{flushleft}